# Hydrogen "penta-graphene-like" structure stabilized by hafnium: a high-temperature conventional superconductor

Hui Xie[1], Yansun Yao[2], Xiaolei Feng[3,4], Defang Duan[1,5,*], Hao Song[1], Zihan Zhang[1], Shuqing Jiang[1], Simon A. T. Redfern[6], Vladimir Z. Kresin[7], Chris J. Pickard[5,8,*] and Tian Cui[9,1,*]

*[1]State Key Laboratory of Superhard Materials, college of Physics, Jilin University, Changchun 130012, China*
*[2] Department of Physics and Engineering Physics, University of Saskatchewan, Saskatoon, Saskatchewan S7N 5E2, Canada*
*[3]Center for High Pressure Science and Technology Advanced Research, Beijing 100094, China*
*[4]Department of Earth Science, University of Cambridge, Downing Site, Cambridge CB2 3EQ, United Kingdom*
*[5]Department of Materials Science & Metallurgy, University of Cambridge, 27 Charles Babbage Road, Cambridge CB3 0FS, United Kingdom*
*[6]Asian School of the Environment, Nanyang Technological University, Singapore 639798*
*[7]Lawrence Berkeley Laboratory, University of California at Berkeley, Berkeley, CA 94720, USA*
*[8]Advanced Institute for Materials Research, Tohoku University 2-1-1 Katahira, Aoba, Sendai, 980-8577, Japan*
*[9]School of Physical Science and Technology, Ningbo University, Ningbo, 315211, People's Republic of China*

[*]Corresponding authors email: duandf@jlu.edu.cn, cjp20@cam.ac.uk, cuitian@jlu.edu.cn

# Abstract

The recent discovery of $H_3S$ and $LaH_{10}$ superconductors with record high superconducting transition temperatures, $T_c$, at high pressure, has fueled the search for room-temperature superconductivity in the compressed superhydrides. Here we predict the existence of an unprecedented hexagonal $HfH_{10}$, with an extraordinarily high $T_c$ of around 213-234 K at 250 GPa. In $HfH_{10}$, the H atoms are arranged in clusters to form a planar "penta-graphene-like" sublattice, in contrast to the covalent sixfold cubic structure in $H_3S$ and clathrate-like structure in $LaH_{10}$. The Hf atom acts as a "precompressor" and electron donor to the hydrogen sublattice. This "penta-graphene-like" $H_{10}$ structure is also found in $ZrH_{10}$, $ScH_{10}$ and $LuH_{10}$ at high pressure, each material showing a high $T_c$ ranging from 134 to 220 kelvin. Our study of dense superhydrides with "penta-graphene-like" layered structures opens the door to the exploration and exploitation of a new class of high $T_c$ superconductors.

# Main

The development of room temperature superconductors is the ultimate goal for superconductivity research. The Bardeen-Cooper-Schrieffer (BCS) theory suggests that metallic hydrogen is likely to be a good candidate for attaining high-temperature superconductivity, due to its high Debye temperature and strong electron-phonon coupling[1-2]. Conceptually, metallization can be achieved in pure hydrogen by dissociating the $H_2$ molecules under extreme conditions. Evidence for molecular dissociation has been found in high-pressure phases of solid hydrogen, where the vibron frequency decreases rapidly with increasing pressure[3], indicating the weakening of the intramolecular bonding along the way to transition to the metallic state. However, although there have been extensive attempts to synthesize metallic hydrogen[4-5], there is a lack of consensus on reported "successes", and superconductivity has not been measured in any of the reported phases.

In the face of tremendous difficulties in the synthesis of pure metallic hydrogen, many investigators have suggested that metallization can be achieved through an alternative approach, by 'precompressing' hydrogen species in a matrix of another element[6], forming a compound. Specifically, highly compressed hydrogen-dominant hydrides are predicted to be able to attain a metallic state and may, therefore, exhibit high $T_c$ superconductivity due to such 'chemical precompression' since the charge density required for metallization can be achieved at lower pressure than that required for pure hydrogen. Recently, the search for high-$T_c$ superconductivity has been expanded from the known species[7-9] to hitherto unknown hydrides, through high-pressure synthesis following theoretical predictions[10-23]. Remarkably, two of the predicted hydrides, $H_3S$[11-13] and $LaH_{10}$[17-19], have been successfully synthesized recently and exhibit record high $T_c$ above 200 K. $H_3S$ has a *bcc* lattice of S with H atoms located halfway between the S atoms, exhibiting three-dimensional covalent metallic characteristics. $LaH_{10}$ has a three-dimensional clathrate-like structure of H with La atoms filling the clathrate cavities, and has been described as an extended metallic hydrogen host structure stabilized by the guest electron donor (La).

By analyzing the superconducting properties of a large number of hydrides, we have summarized four criteria for finding high temperature superconductors in highly compressed hydrides: (i) high symmetry crystal structure, (ii) absence of $H_2$-like molecular units, (iii) a large H contribution to the total electronic density of states (DOS) at the Fermi level, and (iv) strong coupling of electrons on the Fermi surface with high frequency phonons. The search for high-$T_c$ materials has been focused on two families of binary hydrides [see Table S1], covalent sixfold cubic structure[10-14] found in $H_3S$ and $H_3Se$, and clathrate structure[15-22] found in rare earth hydrides $REH_{10}$, $REH_9$ and $REH_6$. These two high-$T_c$ families satisfy the mentioned four criteria, with a common feature that they both adopt three-dimensional sublattices of hydrogen. On the other hand, pure solid hydrogen features pronounced layer-like characters. A prominent example is the phase IV of hydrogen that consists of strongly bonded $H_2$ molecules and weakly bonded graphene-like sheets[24-26]. This structure is considered an important intermediate between the molecular (insulating) and atomic (metallic) crystalline phases of hydrogen. Many hydrides have been predicted to have a layered structure, but their $T_c$ are not very high, *e.g.* $FeH_5$[27], $TeH_4$[28], and $KH_6$[29], etc. One can raise the following question: is there a two-dimensional structure of hydrogen-rich materials at high pressure that can achieve high $T_c$?

To answer this question, we performed an extensive material searching at high-pressure and discovered that a layered hexagonal hafnium decahydride ($HfH_{10}$) indeed has an extraordinarily high $T_c$. In $HfH_{10}$, the H atoms form planar clusters of three $H_5$ pentagons, analogous to penta-graphene, while Hf atoms are intercalated between the clusters on the same plane. A new class of superhydrides, $MH_{10}$ (M = Zr, Sc and Lu), which are isostructural with $HfH_{10}$, have also been predicted to be stable or metastable at high pressure, and all are predicted to possess high superconducting $T_c$.

## Structures of new hydrides and their stabilities

Our main structure searching results for Hf-H system at high pressure are depicted in the convex hull diagrams of Fig. 1 and S1. Considering the non-negligible quantum effects associated with hydrogen, the zero-point energy (ZPE) was calculated and included in the calculation of the formation enthalpies of predicted Hf-H compounds. We analyzed the stability of these compounds with respect to the elemental hafnium and hydrogen end members. As shown in Fig. 1(a), these searches revealed five stable stoichiometries at various pressures, *e.g.*, HfH, $HfH_3$, $HfH_4$, $HfH_6$ and $HfH_{10}$, in addition to the previously known $HfH_2$ and $Hf_4H_{15}$[30-31]. The crystal structures of the predicted compounds are illustrated in Figs. 1(b) and S2. In $HfH_x$ ($x$ = 1 to 4), all hydrogen atoms are present in atomic form. In $HfH_6$, only three hydrogen atoms are in atomic form while the other three form an $H_3$ unit. For $HfH_{10}$, there are two energetically competitive structures, one adopting a $P6_3/mmc$ structure and the other in $P\bar{1}$ symmetry. At 300 GPa, The $P\bar{1}$ structure is calculated to be more enthalpically favorable than the $P6_3/mmc$ structure, but the latter is slightly more stable once the ZPE is included, with the energy difference within 2 meV/atom. It is therefore argued here that the two structures are both valid structures for $HfH_{10}$. The $P\bar{1}$ structure consists of diatomic hydrogen pairs similar to $H_2$ molecules, and this violates criteria (i) and (ii) for attaining high $T_c$. Intriguingly, the $P6_3/mmc$ is a layered structure in which the Hf and H atoms are situated on the same plane [Fig. 1(b) and (c)]. The Hf atoms form a hexagonal sublattice interspersed by H atoms in unique '$H_5$' pentagons akin to the geometry of penta-graphene: three pentagons are fused by edge-sharing to form a $H_{10}$ unit, as depicted in Fig. 1(c). In the proceeding discussion, the superhydride $P6_3/mmc$-$HfH_{10}$ is the main focus because it satisfies both criteria (i) high crystal symmetry and (ii) absence of $H_2$ molecular unit.

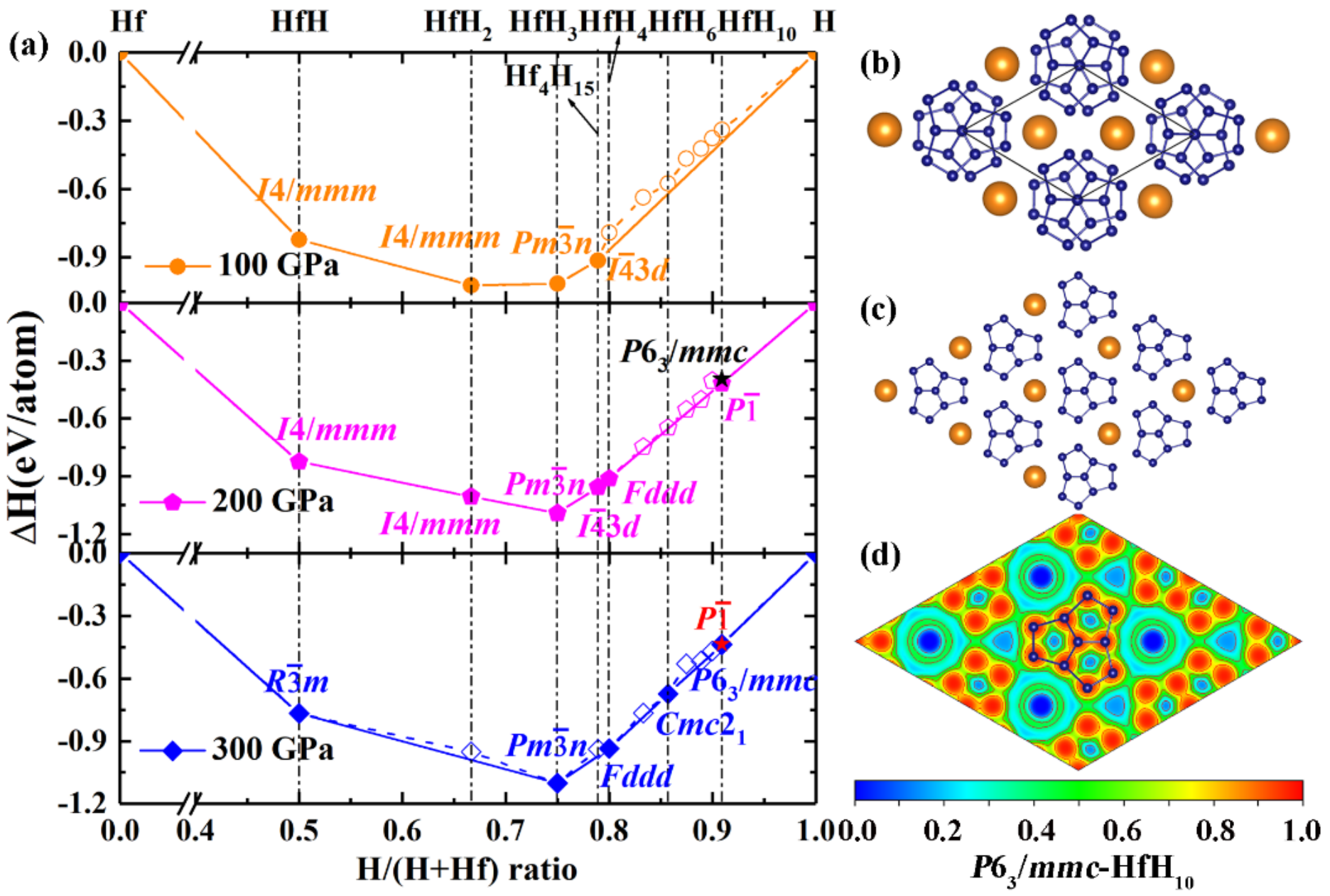

**Fig. 1 | Convex hull of H-rich hafnium hydrides and structure motif of $P6_3/mmc$-$HfH_{10}$.** (a) Formation enthalpies of predicted $HfH_x$ ($x$=1-10) selected from structure searches up to $HfH_{24}$ [see supplementary material], including ZPEs with respect to decomposition into Hf and H under pressure. The $Im\bar{3}m$ structure for hafnium[32], the $P6_3/mc$, $C2/c$ and $Cmca$-12 structures for hydrogen[25] were adopted. Open symbols represent unstable configurations with respect to mixing lines on the convex hull, while solid symbols on the convex hull represent stable configurations. The asterisk near convex hull represents the metastable structure. (b) The crystal structure of $P6_3/mmc$ structure in $HfH_{10}$, (c) a

single (001) plane, (d) ELF on the (001) plane. Golden (large) and small (blue) spheres represent Hf and H atoms, respectively.

We examine the chemical bonding of layered $HfH_{10}$ by analysing electron localization function (ELF), crystal orbital Hamiltonian population (COHP) and Bader charges. As shown in Figs. 1(d) and S3, there is no charge localization between Hf and H, indicating the Hf-H bonding is purely ionic. The ELF values for the $H_{10}$ unit range between 0.6-0.8, showing the evidence for H-H covalent bonding. As depicted in Fig. S4, the calculated COHP show that most of the states below the Fermi level correspond to H-H bonding for $HfH_{10}$, supporting the existence of H-H covalent bonds within the $H_{10}$ unit. Furthermore, Bader charge analysis reveals that the planar $H_{10}$ unit accepts an amount of charge, *e. g.*, ~0.13 $e^-$ per H atom, from nearby Hf atoms, which results in longer H-H distances compared to that in free $H_2$ molecule[33]. The additional electrons reside in the H-H antibonding orbital ($\sigma^*$) and therefore weaken the H-H bonding, thus increasing the H-projected density of states (H-PDOS) at the Fermi level ($\varepsilon_f$).

To further explore this new form of hydrogen, we compared the H-H distances in $HfH_{10}$ with those in $LaH_{10}$[17], atomic H[2] and layered phase-IV of H (*Pc*-48)[24] structures. As shown in Fig. S5, the longest nearest-neighbor H-H distance ($d_{H1\text{-}H2}$) in $HfH_{10}$ is close to the shortest H-H distance in $LaH_{10}$ at the same pressure. The distance between H1 and H2 (second nearest neighbor) atoms approaches that of the atomic structure of hydrogen near 200 GPa. As the pressure increases, the shortest H-H bond length ($d_{H2\text{-}H3}$) gets progressively closer to the H-H distances of *Pc*-48 H ($d_2$). Therefore, this "penta-graphene-like" hydrogen sublattice lies somewhere between atomic and layered hydrogen.

The discovery of apparently stable or metastable structures of $HfH_{10}$ prompt us to further study the Zr-H system at 300 GPa [see Fig. S6]. As might be anticipated, $ZrH_{10}$ adopts the same two competitive structures $P\bar{1}$ and $P6_3/mmc$ as $HfH_{10}$. For high-pressure synthesis, which usually also involves high temperatures, the experimentally realized materials often represent the metastable phases[34]. The recently discovered high-$T_c$ superconducting $LaH_{10}$[18-19], for example, is a metastable compound lying above the convex hull that was previously predicted by Peng et al.[16].

## Evaluation of spectra and $T_c$ for "penta-graphene-like" structure and pressure intervals

According to the criterion (iii) for high $T_c$ superconductivity, a large H contribution to the total electronic DOS at the Fermi level is a critical factor for the development of exceptional superconducting properties. To this end, the projected electronic DOS of $P6_3/mmc$ in $HfH_{10}$ and $ZrH_{10}$ were calculated and are illustrated in Fig. 2. One can see that both phases are metallic with a large total electronic DOS and significant hydrogen contribution to the electronic DOS at the Fermi level. Remarkably, the electronic DOS exhibits van Hove singularities near the Fermi level, indicating a large electron-phonon coupling (EPC) strength bound up with hydrogen phonon modes. The total DOS of $H_3S$, $LaH_{10}$, $HfH_{10}$ and $ZrH_{10}$ at 300 GPa are compared in Fig. 2(c). It is shown that $HfH_{10}$ and $LaH_{10}$ have a comparable electronic DOS at the Fermi level, and these are notably larger than those of $ZrH_{10}$ and $H_3S$.

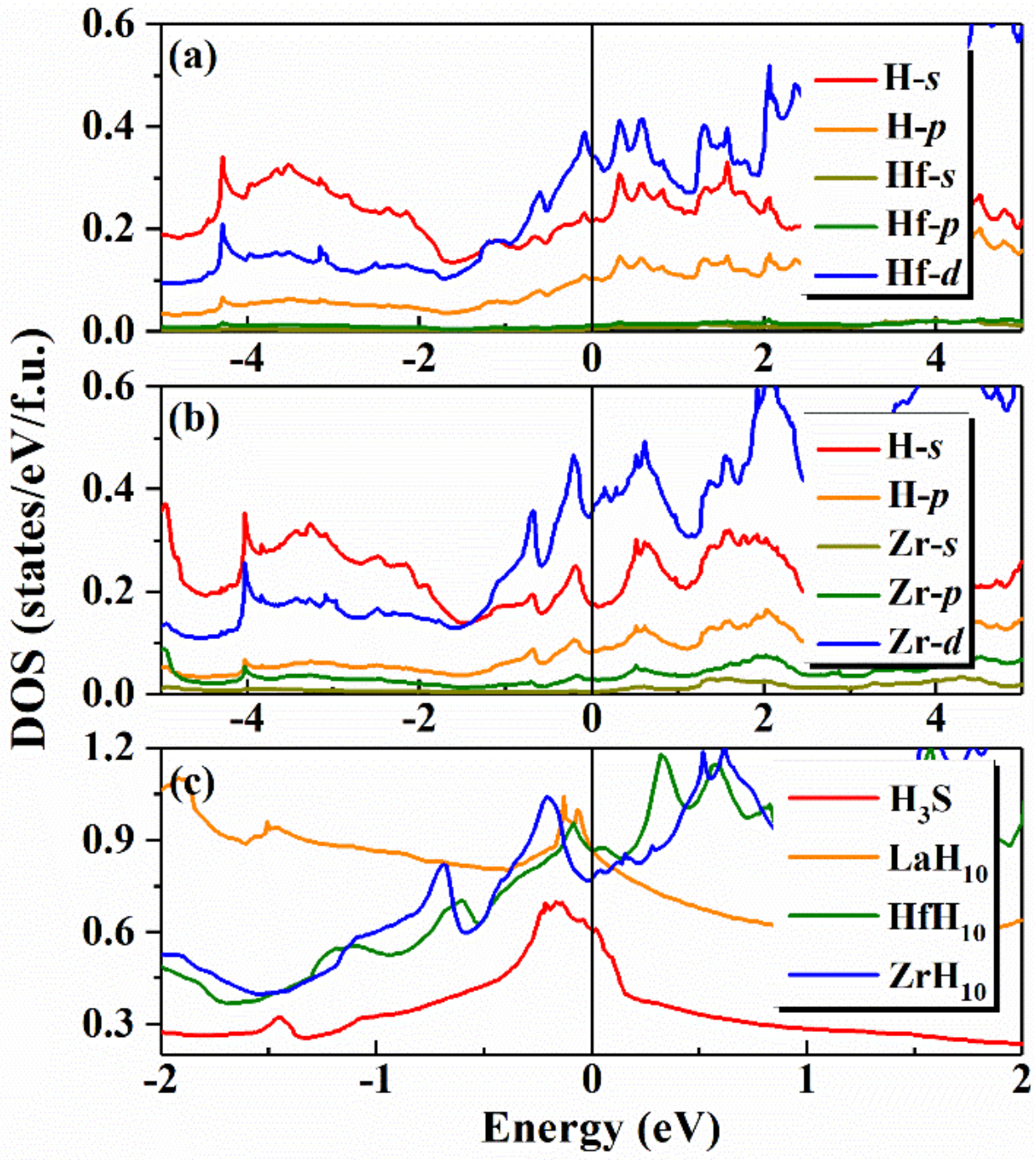

**Fig. 2 | Calculated electronic DOS at 300 GPa.** The projected electronic DOS of (a) $P6_3/mmc$-$HfH_{10}$ and (b) $P6_3/mmc$-$ZrH_{10}$. (c) The total electronic DOS of $H_3S$, $LaH_{10}$, $HfH_{10}$ and $ZrH_{10}$ around the van Hove singularities.

To examine the superconductivity in the "penta-graphene-like" structure, we calculate the average phonon frequency and EPC as shown in Table S4. For $HfH_{10}$, our EPC calculation yields a large λ of 2.16 at 300 GPa which is benefited from large H-PDOS and high frequency vibrations (above 10 THz) due to hydrogen which contribute 70 % to the value of $\lambda$ [Fig. 3(a)]. It is obvious that the large λ satisfy the last criterion (iv). Since λ is larger than 1.5, we calculated $T_c$ using three approaches: Allen-Dynes modified McMillian equation (A-D) [Eq. S12][35], Matsubara-type linearized Eliashberg equation (LE)[21], and Gor'kov-Kresin equation (G-K)[36], all of which were designed to estimate the $T_c$ for materials with strong electron-phonon coupling. The results show that $HfH_{10}$ is an extraordinary superconductor with a $T_c$ of 151-166 K (A-D), 214-228 K (LE), and 197-220 K (G-K) using $\mu^* = 0.1$-0.13 at 300 GPa. To narrow down the range of $T_c$, we calculated the $T_c$ of $H_3S$ and $LaH_{10}$ at 200 GPa and compared them to the experimental values to obtain appropriate parameters for this family of materials. As presented in Fig. S7, the calculated $T_c$ with $\mu^* = 0.13$ using G-K and LE equations are close to the experimental values, while those estimated by A-D equation are much lower. In the following, we will estimate the $T_c$'s using the G-K equation. With the pressure decreased to 250 GPa, λ and $T_c$ for $HfH_{10}$ increase to 2.77 and 213-234 K with $\mu^* = 0.1$-0.13, respectively; these values are higher than those for $H_3S$. For $ZrH_{10}$, a large EPC parameter $\lambda$ of 1.59 is calculated at 300 GPa, of which 74% is due to contributions by H atoms [Fig. 3(b)]. A high superconducting transition temperature $T_c$ of 194-218 K is therefore estimated for $ZrH_{10}$. At 250 GPa, $T_c$ increases to 199-220 K with stronger $\lambda$ of 1.77. We also calculated the electronic DOS and $T_c$ of $P\bar{1}$-$HfH_{10}$ at 200 GPa [see Fig. S8 and Table S6]. It is found that, as expected, the existence of $H_2$ units reduce the electronic DOS at the Fermi energy, and a low EPC parameter of 0.72, thereby limiting its superconductivity with $T_c$ of 28.9-37.4 K ($\mu^* = 0.1$-0.13). Therefore, the four criteria for superconductivity provide important guidance in the search for high-temperature superconductors in compressed superhydrides.

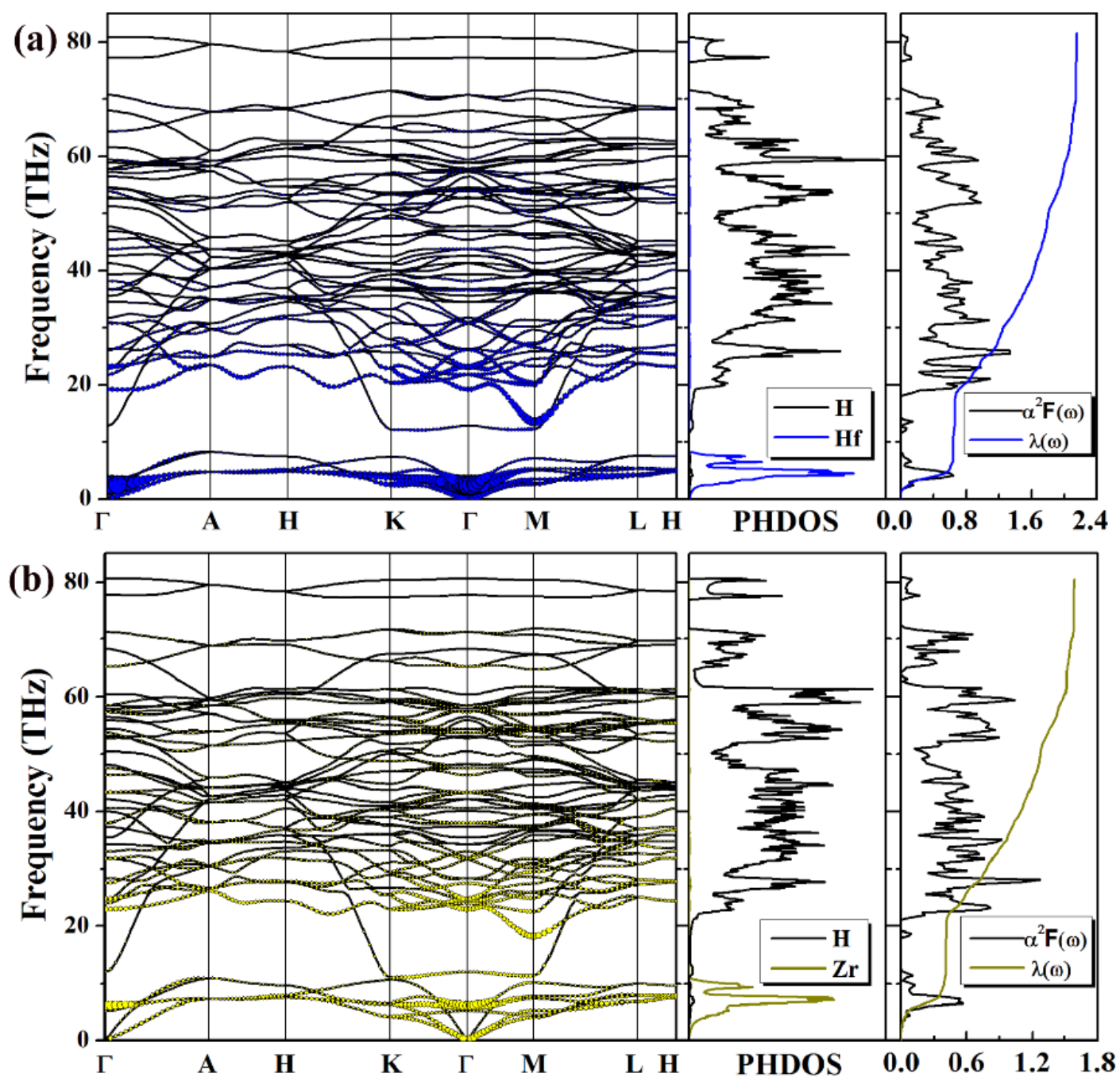

**Fig. 3 | Phonon properties and Eliashberg spectral function for $HfH_{10}$ and $ZrH_{10}$ with *P*6₃/*mmc* structure.** Phonon dispersion curves (left), density of states (middle) and Eliashberg spectral function $\alpha^2F(\omega)$ together with the electron-phonon integral $\lambda(\omega)$ (right) of (a) $HfH_{10}$ and (b) $ZrH_{10}$ at 300 GPa. The size of the solid dot on phonon spectra signifies the contribution to electron-phonon coupling.

To further understand the superconductivity of the 'clathrate-like' and "penta-graphene like" decahydrides, we compared the calculated $T_c$'s and essential parameters for $YH_{10}$, $LaH_{10}$, $HfH_{10}$ and $ZrH_{10}$ at 300 GPa, illustrated in Fig. 4. We find that $T_c$ decreases in the order of $YH_{10} > LaH_{10} > HfH_{10} > ZrH_{10}$, consistent with the decreasing trend of optical superconducting transition temperature caused by the interaction of electrons with optical phonons ($T_c^0$), the contribution of H-PDOS to the total DOS at the Fermi level ($P_H$) and optical electron-phonon coupling ($\lambda_{opt}$). Thus, the high $T_c$ in these hydrides is mainly attributed to the interaction of electrons with optical phonons and high DOS at the Fermi level associated with H atoms, in agreement with both superconducting criteria (iii) and (iv).

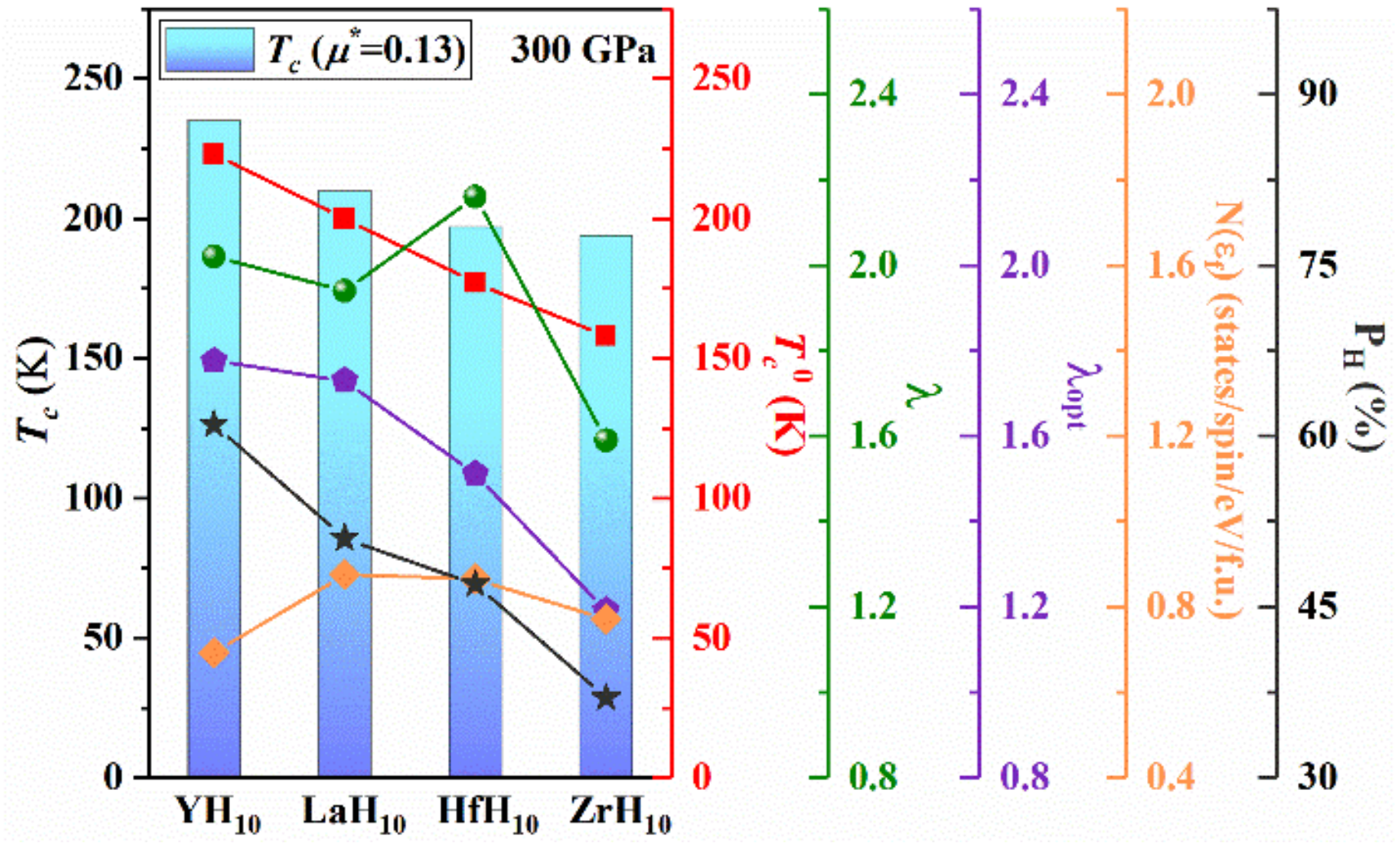

**Fig. 4 | Calculated superconducting parameters for $YH_{10}$, $LaH_{10}$, $HfH_{10}$ and $ZrH_{10}$ at 300 GPa.** The superconducting critical temperature using G-K equation with $\mu^* = 0.13$ ($T_c$), the critical temperature caused by the interaction of electrons with optical phonons ($T_c^0$), EPC parameter ($\lambda$),strength of the interaction of electrons with optical phonons ($\lambda_{opt}$), electronic DOS at the Fermi level $N(\varepsilon_f)$ and the contribution of H atoms DOS to the total DOS at the Fermi energy ($P_H$).

The fact that "penta-graphene like" $HfH_{10}$ and $ZrH_{10}$ have very high $T_c$ naturally raises the question: can any other hydrides adopt the same structure and display high $T_c$ as well? The elements of Hf and Zr have three features in common: similar Pauling electronegativity (~1.3), similar atomic radius (~1.6 Å), and *d* states in valence subshells. After searching the periodic table of elements, we found that Mg, Sc, Lu and Th have some similar properties. Phonon dispersion relations were calculated to test the stability of the corresponding decahydrides. The calculations did confirm that both $ScH_{10}$ and $LuH_{10}$ are dynamically stable, while $MgH_{10}$ and $ThH_{10}$ are not. For $MgH_{10}$ [Fig. S9] its dynamic instability could be due to the absence of *d* electrons in Mg and insufficient number of electrons transferred to H. For $ThH_{10}$ [Fig. S9], the large atomic radius of Th (1.8 Å) may be responsible for its dynamic instability. We note that the hexagonal structure $P6_3/mmc$ was first noted by Peng et al.[16] for $ScH_{10}$. It was not discussed in depth, the main focus of this work being on the remarkable clathrate structure. Later, an orthogonal structure *Cmcm* in $ScH_{10}$ was predicted[37]. We thus calculate the enthalpies of *Cmcm* and $P6_3/mmc$ phases as a function of pressure and find that these phases are indeed competitive [Fig. S10]. The electron-phonon coupling calculation for the $P6_3/mmc$- $ScH_{10}$ yields a $\lambda$ of 1.16 and $T_c$ ranging 134-158 K with $\mu^*$ = 0.1-0.13 at 250 GPa [Figs. S11-S12]. For Lu-H system, we perform extensive structure searching and find that $LuH_{10}$ with $P6_3/mmc$ phase is stable at 300 GPa [Fig. S13]. Phonon calculation shows that it is dynamical stable down to at least 200 GPa [Fig. S14]. At 200 GPa, $LuH_{10}$ is found to be a good superconductor with a relatively high $T_c$ of 134-152 K, comparable with to most other lanthanide hydrides. Although the emergence of 5*d* electron in Lu suppresses the contribution of 4*f* electrons and improves the contribution of hydrogen atoms at the Fermi level [Fig. S15], the relatively low total DOS at the Fermi level limits to some extent its superconductivity.

## Isotope effect

Since the substitution of deuterium for hydrogen (H→D) affects the optical modes only, whereas the value of $T_c$ is affected by both acoustic and optical modes, the value of the isotope coefficient and its closeness to the $\alpha_{max}$ = 0.5 reflects the relative impact of the high frequency optical modes on $T_c$ and the interplay of the optical and acoustic modes. It is worth mentioning that the value of $T_c$ would be reduced upon H→D substitution, since the high frequency hydrogen modes determine the value of the critical temperature in high-$T_c$ hydrides. Thus, the isotope coefficient ($\alpha$) was calculated to estimate the $T_c$ of $MD_{10}$ ($T_c^D$). As shown in Table S5, the coefficients are 0.42-0.43 for $HfH_{10}$ at 250 GPa, 0.38-0.39 for $ZrH_{10}$ at 250 GPa, 0.37-0.38 for $ScH_{10}$ at 250 GPa, and 0.44-0.45 for $LuH_{10}$ at 200 GPa with $\mu^*$ = 0.1-0.13, respectively. The $\alpha$ values in these phases are relatively large, suggesting that the pairing, particularly in $HfH_{10}$ and $LuH_{10}$, is dominated by the optical H modes, which yields the relatively low $T_c^D$. The $T_c^D$ was described according to the equation of $T_c/T_c^D = (M_H/M_D)^{-\alpha}$, where $T_c$ is obtained from the G-K equation. In our cases, $T_c^D$ values are 159-174 K for $HfD_{10}$, 153-168 K for $ZrD_{10}$, 104-121 K for $ScD_{10}$, and 99-111 K for $LuD_{10}$, respectively, providing a reference for future experiments.

## New materials

Our extensive first-principles structure searches have revealed the appearance of stable or metastable "penta-graphene like" clustered $H_{10}$ structure in $HfH_{10}$, $ZrH_{10}$, $ScH_{10}$ and $LuH_{10}$. Electronegativity, atomic radius, and valence configuration of the metal element are all found to play critical roles in the stabilization of this "penta-graphene like" hydrogen sublattice, and fine-tune the superconductivity of these materials. We want to pay a special attention to $HfH_{10}$ which is predicted to be a high temperature superconductor with estimated $T_c$ of 213-234 K at 250 GPa. It is the first example of two-dimensional structure in hydrogen-rich materials exhibiting high $T_c$ above 200 K, which will encourage scientists to search for high-temperature superconductors in layered hydrogen-rich materials. It is also the hydride that has the highest $T_c$ to date in the transition metal hydrides. This "penta-graphene like" clustered $H_{10}$ structure is a structural model for superconducting hydrides with $T_c$ higher than 200 K which is different from the covalent bonded H structure in $H_3S$ and clathrate H structure in $LaH_{10}$. One can state that at present there are three different model structures for high $T_c$ hydrides. The new structure model described in this paper further confirmed four criteria for high $T_c$ superconductivity in hydrogen-rich material proposed in the beginning of the paper: (i) high symmetry is manifested by hexagonal structure $P6_3/mmc$, (ii) absence of $H_2$-like molecular units; indeed, three $H_5$ pentagons fused a planar $H_{10}$ clusters with the H-H distances which are between the values for the metallic phase and phase IV of solid hydrogen, (iii) a large H contribution to the total electronic density of states (DOS) at the Fermi level, which in this case is nearly one half, and (iv) strong coupling of electrons on the Fermi surface with high frequency phonons with large value of the coupling constant equal to 2.77. The emergence of "penta-graphene like" clustered $H_{10}$ structure is quite significant, which will stimulate further study of the hydrogen-based family of new superconducting materials and certainly help to develop this promising field. The formulated four criteria for high $T_c$ superconductivity provide a rationale for searching room-temperature superconductors in ternary or quaternary hydrogen-rich materials in the future, which have not been well explored to date.

# Methods

We carried out extensive structure searches using Ab Initio Random Structure Searching (AIRSS), Crystal structure AnaLYsis by Particle Swarm Optimization (CALYPSO) and Universal Structure Predictor: Evolutionary Xtallography (USPEX)[38-40] at pressures from 100 to 300 GPa, during which at least 15,000 structures were generated. Geometrical optimization calculations for Hf-H and Zr-H systems were performed using the on-the-fly generation of ultrasoft pseudopotentials with a plane-wave basis set energy cutoff of 800 eV, as implemented in CASTEP (Cambridge Sequential Total

Energy Package) code[41]. In $ScH_{10}$ and Lu-H cases, the all-electron projector-augmented wave method (PAW)[42] pseudopotentials with a cutoff energy of 1000 eV were employed using VASP (Vienna Ab initio Simulation Package) code[43]. The calculations were carried out with the Perdew-Burke-Ernzerhof generalized gradient approximation[44] density functional, except for the phonon calculation of Zr-H phases with the GGA_PBEsol[45] functional. The Brillouin zone was sampled using a k-point mesh of $2\pi \times 0.03$ Å$^{-1}$ for structure relaxations, and a denser spacing of $2\pi \times 0.02$ Å$^{-1}$ for electronic property calculations. Phonon frequencies and zero point energies were calculated using the supercell approach[46], as implemented in PHONOPY[47] and CASTEP codes. The electron-phonon coupling parameters were computed within linear response theory with the Quantum ESPRESSO package[48]; see the Supporting Information for further details.

## Acknowledgements

We thank Professor Yanming Ma and Dr. Dmitrii V. Semenok for many interesting and stimulating discussions. This work was supported by the National Key R&D Program of China (No. 2018YFA0305900), National Natural Science Foundation of China (Nos. 51632002 and 11674122), Program for Changjiang Scholars and Innovative Research Team in University (No. IRT_15R23), the 111 Project (No. B12011), and the Natural Sciences and Engineering Research Council of Canada (NSERC). C.J.P. acknowledges financial support from the Engineering and Physical Sciences Research Council (Grant EP/P022596/1) and a Royal Society Wolfson Research Merit award. X. F. acknowledges China Scholarship Council. Parts of the calculations were performed in the High Performance Computing Center (HPCC) of Jilin University and TianHe-1(A) at the National Supercomputer Center in Tianjin.

## Author Contributions

D.D. and T.C. designed research; H.X., D.D. and C.J.P performed calculations; H.X., D.D., C.J.P., V.Z.K., X.F., H.S., S.J., Y.Y. and T.C. analyzed data; H.X., Y.Y., X.F., D.D., Z.Z., S.A.T.R., V.Z.K., C.J.P., and T.C. wrote and revised the paper.

# Competing interests

The authors declare no competing interests

# Supporting Information
# For
## Hydrogen "penta-graphene-like" structure stabilized by hafnium: high-temperature conventional superconductor

Hui Xie[1], Yansun Yao[2], Xiaolei Feng[3,4], Defang Duan[1,5,*], Hao Song[1], Zihan Zhang[1], Shuqing Jiang[1], Simon A. T. Redfern[6], Vladimir Z. Kresin[7], Chris J. Pickard[5,8,*] and Tian Cui[9,1,*]

*[1]State Key Laboratory of Superhard Materials, college of Physics, Jilin University, Changchun 130012, China*
*[2] Department of Physics and Engineering Physics, University of Saskatchewan, Saskatoon, Saskatchewan S7N 5E2, Canada*
*[3]Center for High Pressure Science and Technology Advanced Research, Beijing 100094, China*
*[4]Department of Earth Science, University of Cambridge, Downing Site, Cambridge CB2 3EQ, United Kingdom*
*[5]Department of Materials Science & Metallurgy, University of Cambridge, 27 Charles Babbage Road, Cambridge CB3 0FS, United Kingdom*
*[6]Asian School of the Environment, Nanyang Technological University, Singapore 639798*
*[7]awrence Berkeley Laboratory, University of California at Berkeley, Berkeley, CA 94720, USA*
*[8]Advanced Institute for Materials Research, Tohoku University 2-1-1 Katahira, Aoba, Sendai, 980-8577, Japan*
*[9]School of Physical Science and Technology, Ningbo University, Ningbo, 315211, People's Republic of China*

[*]Corresponding authors email: duandf@jlu.edu.cn, cjp20@cam.ac.uk, cuitian@jlu.edu.cn

# Content

# Computational details

High-pressure structural predictions within *ab initio* calculations are implemented with the AIRSS (Ab Initio Random Structure Searching) code[1], which effectiveness has been confirmed by the successful applications to discovering the structures of solids, point defects, surfaces, and clusters. CALYPSO (Crystal structure AnaLYsis by Particle Swarm Optimization)[2-3] and USPEX (Universal Structure Predictor: Evolutionary Xtallography)[4-6] codes were also used for searching high-pressure structures of Zr-H system, giving the same results as that generated by using AIRSS.

Given that most of the recently discovered superconducting superhydrides are stable at pressures above 100 GPa, some even at pressures higher than 300 GPa, we set our initial structure searches at 300 GPa. Extensive random structure searches were carried out for 40 compositions which resulted in the generation of approximately 15,000 structures, among which $HfH_{24}$ showed the highest hydrogen content. A convex hull was constructed for each predicted compounds and those laying on or within ~ 0.1 eV/atom above the convex hull were selected for refined structural optimization with higher criteria (Fig. S1). As shown in Fig. S1 (a), the structure searches have identified an extremely H-rich species of $HfH_{18}$ with general accuracy, while it loses its stability in high precision calculations, as shown in Fig S1 (b). For all thermodynamically stable stoichiometries (those on the convex hull), $HfH_{10}$ showed the highest hydrogen content. Subsequently, fixed-composition searches were carried out restricting the Hf:H ratio to (or below) 1:10, *i.e.*, $HfH_x$ ($x$=1-10), and these were performed at 100, 200 and 300 GPa.

For Hf-H and Zr-H systems, structural relaxations were performed by the on-the-fly (OTF) generation of ultrasoft pseudopotentials in CASTEP (Cambridge Sequential Total Energy Package) code[7], where the valence electrons configurations are $4f^{14}5s^25p^65d^26s^2$ for Hf, $4s^24p^64d^25s^2$ for Zr and $1s^1$ for H. The cutoff energy was chosen to be 800 eV. In other cases, the all-electron projector-augmented wave method (PAW)[8] pseudopotentials were employed using VASP code[9] with cutoff energy of 1000 eV. The valence electrons of the pseudopotentials are $3p^63d^14s^2$ for Sc, $4f^{14}5s^25p^65d^16s^2$ for Lu, $2s^22p^63s^2$ for Mg, $6p^66d^27s^2$ for Th and $1s^1$ for H. The exchange-correlation functional was described using Perdew–Burke–Ernzerhof (PBE) of generalized gradient approximation (GGA)[10-11] for all systems. A Monkhorst-Pack[12] k-point mesh of $2\pi \times 0.03$ Å$^{-1}$ was used to ensure that the enthalpy calculations converged effectively to within less than 1 meV/atom.

All-electron full-potential linearized augmented plane wave (FP-LAPW) method with WIEN2k code[13] was performed to test the validity of the PAW pseudopotentials used in VASP and OTF pseudopotentials used in CASTEP. In the full-potential calculations, the FP-LAPW basis function of RKmax = 5 and 3000 k points in the electronic integration of the Brillouin zone were used to achieve a satisfactory degree of convergence. The Perdew-Burke-Ernzerhof generalized gradient approximation (GGA_PBE) exchange correlation functional was chosen. The pressure-volume curves of $MH_3$ were fitted by third-order Birch-Murnaghan (BM) equation[14].

Electronic properties were calculated by means of the VASP code using a k-point mesh of $2\pi \times 0.02$ Å$^{-1}$. Electron localization function (ELF)[15] was computed to describe and visualize chemical bonds in multielectron systems, which can easily reveal atomic shell structure and core, binding, and lone electron pairs. The ELF value is in the range of 0-1. The upper limit ELF = 1 corresponds to perfect localization and the value ELF = 0.5 corresponds to electron gas-like pair probability. The crystal orbital Hamiltonian population (COHP) and integrated COHP (ICOHP) were also calculated using LOBSTER[16], which is commonly used for differentiating covalent and non-covalent bonding in chemistry. And Bader charge analysis[17-18] was performed to determine charge transfer.

The phonon and zero point energy calculations were carried out using the PHONOPY[19] and CASTEP codes. In the calculations, GGA_PBE and GGA_PBEsol[20] exchange correlation functional were chosen for Hf-H and Zr-H systems, respectively. The linear response theory was performed to calculate the electron-phonon coupling (EPC) with the Quantum ESPRESSO package[21]. Convergence tests gave us an appropriate kinetic energy cutoff of 90 Ry. Pseudopotentials were generated by a Troullier-Martins norm-conserving scheme[22] which is found to be particular efficient for systems containing first-row elements and transition metals. Self-consistent electron density and electron-phonon coupling were calculated by employing 12×12×18 k-mesh and 4×4×6 q-mesh for $P6_3/mmc$-$MH_{10}$ (M = Hf, Zr, Sc and Lu). The superconducting transition temperatures $T_c$ were calculated using three different equations as follows[23-26].

# Equations for calculating $T_c$ and related parameters

In this section we will introduce the equations for calculating $T_c$ of hydrides. The metallic hydrides which possess high $T_c$ were considered as conventional superconductors (the pairing mechanism is the electron-phonon interaction). And the high $T_c$ is mainly attributed to the motion of H atoms. In order to study influence of the forms (cages, penta-graphene-like and covalent compound) of H -structures in hydrides on $T_c$, we analyse as the dominant the contribution of optical modes, because of large difference in masses. Indeed, as we know, the optical modes are determined by the hydrogen motion, whereas the acoustic modes correspond to the motion of heavy ion.

The Eliashberg equation for the pairing order parameter $\Delta(\omega_j)$ has the following form (at T = $T_c$)[23]:

$$\Delta(\omega_j)Z = \pi T_c \sum_i \int d\omega \frac{\alpha^2(\omega)F(\omega)}{\omega} \frac{\omega^2}{\omega^2+(\omega_j-\omega_i)^2} \frac{\Delta\omega_i}{|\omega_i|}. \tag{S1}$$

Here, $Z$ is the renormalization factor:

$$Z = 1 + \frac{\pi T_c}{\omega_j} \sum_i \int d\omega \frac{\alpha^2(\omega)F(\omega)}{\omega} \frac{\omega^2}{\omega^2+(\omega_j-\omega_i)^2} \frac{\omega_i}{|\omega_i|}. \tag{S1}$$

And $\omega_i$ is the Matsubara frequency:

$$\omega_i = (2i+1)\pi T_c \ \ (i = 0,\ \pm 1,\ \pm 2, \dots), \tag{S3}$$

In our calculations, the cut off Matsubara frequency is $|i| \le 24$ (when $|i_{max}|$ = 24., $\omega_{max}$ ≈ 300 THz for the $T_c$ ≈ 200 K). The Eliashberg equation is a non-linear integral equation. Here we used the Matsubara-type linearized Eliashberg equations, introduced by Bergmann and Rainer[27-28] and Allen[29], and later developed by Kvashin et al.[30], to get the approximate solution of the Eliashberg equation (LE):

$$\Delta(\omega = \omega_j, T) = \Delta_i(T) = \pi T \sum_i \frac{[\lambda(\omega_j-\omega_i)-\mu*]}{\rho+|\omega_i+\pi T \sum_k (sign\omega_k)\cdot\lambda(\omega_i-\omega_k)|} \cdot \Delta_i(T), \tag{S4}$$

where $\mu^*$ is a Coloumb pseudopotential. The function $\lambda(\omega_j - \omega_i)$ relates to effective electron-electron interaction via exchange of phonons and takes forms:

$$\lambda(\omega_j - \omega_i) = 2\int_0^\infty \frac{\omega\cdot\alpha^2F(\omega)}{\omega^2+(\omega_j-\omega_i)^2}\, d\omega. \tag{S5}$$

Transition temperature $T_c$ can be found as a solution of equation $\rho(T_c) = 0$, where $\rho(T)$ is defined as $max(\rho)$ providing that $\Delta(\omega)$ is not a zero function of $\omega$ at fixed temperature. Here provided a way to calculate $\rho(T)$:

$$\rho(T) = \max\{\text{eigenvalue}(K_{mn})\}\text{e}, \tag{S6}$$

$$K_{mn} = F(m-n) + F(m+n+1) - 2\mu^* - \delta_{mn}[2m+1+F(0)+2\sum_{l=1}^{m} F(l)], \tag{S7}$$

$$F(n) = F(n,T) = 2\int_0^{\omega_{max}} \frac{\alpha^2F(\omega)}{\omega^2+(2\pi\cdot T\cdot n)^2} \cdot \omega d\omega. \tag{S8}$$

Here, the $\rho$(T) = 0 when $T = T_c$.

The EPC spectral function $\alpha^2F(\omega)$ is given by[24]:

$$\alpha^2F(\omega) = \frac{1}{2\pi N(\varepsilon_f)} \sum_{qj} \frac{\gamma_{qj}}{\omega_{qj}} \delta(\omega - \omega_{qj}) w(q). \tag{S9}$$

$\gamma_{qj}$ is the phonon linewidth,

$$\gamma_{qj} = 2\pi\omega_{qj} \sum_{nm} \int \frac{d^3k}{\Omega_{BZ}} \left|g^{j}_{kn,k+qm}\right|^2 \delta(\varepsilon_{kn} - \varepsilon_F)\delta(\varepsilon_{k+qm} - \varepsilon_F), \quad \text{(S10)}$$

where $\Omega_{BZ}$ is the volume of the BZ, $g^{j}_{kn,k+qm}$ are the electron-phonon matrix elements, $\varepsilon_F$ is the Fermi energy, $\varepsilon_{kn}$ and $\varepsilon_{k+qm}$ are the Kohn-Sham eigenvalues in the band $n$ at wave vector $k$ and band $m$ at wave vector $k + q$, respectively.

The Allen-Dynes modified McMillan equation (A-D) which is the approximate analytic solution of the Eliashberg equations valid at $\lambda < 1.5$ is[24]:

$$T_c = \frac{\omega_{log}}{1.2} \exp\left[-\frac{1.04(1+\lambda)}{\lambda-\mu^*(1+0.62\lambda)}\right], \quad \text{(S11)}$$

when $\lambda > 1.5$, we have the following expression containing the corrections $f_1$ and $f_2$ ::

$$T_c = \frac{f_1 f_2 \omega_{log}}{1.2} \exp\left[-\frac{1.04(1+\lambda)}{\lambda-\mu^*(1+0.62\lambda)}\right], \quad \text{(S12)}$$

two separate correction factors $f_1$ and $f_2$ are given by[24]:

$$f_1 = \sqrt[3]{\left[1 + \left(\frac{\lambda}{2.46(1+3.8\mu^*)}\right)^{\frac{3}{2}}\right]}, f_2 = 1 + \frac{(\frac{\bar{\omega}_2}{\omega_{log}}-1)\lambda^2}{\lambda^2+[1.82(1+6.3\mu^*)\frac{\bar{\omega}_2}{\omega_{log}}]^2}, \quad \text{(S13)}$$

Here $\bar{\omega}_2$ is mean square frequency,

$$\bar{\omega}_2 = \sqrt{\frac{2}{\lambda}\int \alpha^2 F(\omega)\, \omega d\omega}\ , \quad \text{(S14)}$$

$\omega_{log}$ is the logarithmic average frequency and $\mu^*$ is the Coulomb pseudopotential, for which we use the widely accepted range of 0.1-0.13. The $\omega_{log}$ and EPC constant $\lambda$ were calculated as:

$$\omega_{log} = \exp\left[\frac{2}{\lambda}\int \frac{d\omega}{\omega}\alpha^2 F(\omega) \ln \omega\right], \quad \text{(S15)}$$

$$\lambda = 2\int \frac{\alpha^2 \mathrm{F}(\omega)}{\omega} d\omega = \sum_{qj} \lambda_{qj} w(q), \quad \text{(S16)}$$

$$\lambda_{qj} = \frac{\gamma_{qj}}{\pi\hbar N(\varepsilon_f){\omega_{qj}}^2}, \quad \text{(S17)}$$

where $\lambda_{qj}$ is the mode EPC parameter[31] and $w(q)$ is the weight of phonon mode $j$ at wave vector $q$ in the first Brillouin zone (BZ).

The phonon spectrum of hydrides contains acoustic and optical modes and is very broad because of the presence of light hydrogen ions. In order to study the impact of H atoms, Gor'kov and Kresin (G-K) introduced the coupling constants $\lambda_{\mathrm{opt}}$ and $\lambda_{\mathrm{ac}}$ describing the interaction of electrons with optical and acoustic phonons, respectively[25-26].

$$\lambda_{\mathrm{ac}} = 2\int_0^{\omega_1} \frac{\alpha^2 F(\omega)}{\omega} d\omega,\ \lambda_{\mathrm{opt}} = 2\int_{\omega_1}^{\omega_m} \frac{\alpha^2 F(\omega)}{\omega} d\omega,\ \lambda_{\mathrm{ac}} + \lambda_{\mathrm{opt}} = \lambda, \quad \text{(S18)}$$

where $\omega_1$ is the maximum frequency for the acoustic modes, $\omega_m$ is the maximum value of frequency of the optical modes. The average values are defined as follows:

$$\tilde{\omega}_{\mathrm{ac}} = \langle {\omega_{\mathrm{ac}}}^2 \rangle^{\frac{1}{2}},\ \langle {\omega_{\mathrm{ac}}}^2 \rangle = \frac{2}{\lambda_{\mathrm{ac}}} \int_0^{\omega_1} d\omega \cdot \omega^2 \frac{\alpha^2 F(\omega)}{\omega} = \frac{2}{\lambda_{\mathrm{ac}}} \int_0^{\omega_1} \alpha^2 F(\omega) \omega d\omega, \quad \text{(S19)}$$

$$\tilde{\omega}_{\mathrm{opt}} = \langle {\omega_{\mathrm{opt}}}^2 \rangle^{\frac{1}{2}},\ \langle {\omega_{\mathrm{opt}}}^2 \rangle = \frac{2}{\lambda_{\mathrm{opt}}} \int_{\omega_1}^{\omega_{\mathrm{m}}} d\omega \cdot \omega^2 \frac{\alpha^2 F(\omega)}{\omega} = \frac{2}{\lambda_{\mathrm{opt}}} \int_{\omega_1}^{\omega_{\mathrm{m}}} \alpha^2 F(\omega) \omega d\omega, \quad \text{(S20)}$$

Then the generalized Eliashberg equation has the form:

$$\Delta(\omega_n) Z = \pi T \sum_{\omega_{n'}} \left[ \lambda_{\mathrm{opt}} \frac{\tilde{\Omega}^2_{\mathrm{opt}}}{\tilde{\Omega}^2_{\mathrm{opt}} + (\omega_n - \omega_{n'})^2} + \lambda_{\mathrm{ac}} \frac{\tilde{\Omega}^2_{\mathrm{ac}}}{\tilde{\Omega}^2_{\mathrm{ac}} + (\omega_n - \omega_{n'})^2} \right] \frac{\Delta(\omega_{n'})}{|\omega_{n'}|}\Bigg|_{T=T_c}, \quad \text{(S21)}$$

For our predicted hydrides the $\lambda_{\mathrm{ac}} << \lambda_{\mathrm{opt}}$, we assume that:

$$T_c = T_c^{opt} + \Delta T_c^{ac},\ \text{and}\ T_c^{\mathrm{opt}} >> \Delta T_c^{\mathrm{ac}} \quad \text{(S22)}$$

As a result, the expression for $T_c$ can be written in the form:

$$T_c = \left[ 1 + 2 \frac{\lambda_{\mathrm{ac}}}{\lambda_{\mathrm{opt}} - \mu^*} \cdot \frac{1}{1 + \eta^{-2}} \right] T_c^0, \eta = \frac{\tilde{\omega}_{\mathrm{ac}}}{\pi T_c^0}, T_c^0 \equiv T_c^{opt}. \quad \text{(S23)}$$

Here the $T_c^0$ is defined as the transition temperatures caused by the interaction of electrons with optical phonons only, for $\lambda_{\mathrm{opt}} \lesssim 1.5$:

$$T_c^0 = \frac{\tilde{\omega}_{\mathrm{opt}}}{1.2} \exp\left[ - \frac{1.04(1 + \lambda_{\mathrm{opt}})}{\lambda_{\mathrm{opt}} - \mu^*(1 + 0.62\lambda_{\mathrm{opt}})} \right]. \quad \text{(S24)}$$

For $\lambda_{\mathrm{opt}} > 1.5$:

$$T_c^0 = \frac{0.25\tilde{\omega}_{\mathrm{opt}}}{[e^{\frac{2}{\lambda_{eff}}} - 1]^{1/2}}. \quad \text{(S25)}$$

Here the $\lambda_{\mathrm{eff}}$ is defined as follows:

$$\lambda_{\mathrm{eff}} = (\lambda_{\mathrm{opt}} - \mu^*)\left[1 + 2\mu^* + \lambda_{\mathrm{opt}} \mu^* t(\lambda_{\mathrm{opt}})\right]^{-1}, \quad \text{(S26)}$$

$$t(x) = 1.5 \exp(-0.28)\, x. \quad \text{(S27)}$$

The values of the isotope coefficient α and superconductive gap were calculated with use of the following expressions:

$$\alpha = \frac{1}{2}\left[1 - 4 \frac{\lambda_{ac}}{\lambda_{opt}} \frac{\eta^2}{(\eta^2 + 1)^2}\right], \quad \text{(S28)}$$

$$\frac{2\Delta(0)}{T_c} = 3.52 \left[ 1 + 5.3 \left( \frac{T_c}{\tilde{\omega}} \right)^2 \ln \frac{\tilde{\omega}}{T_c} \right]. \quad \text{(S29)}$$

# Comparison of high $T_c$ hydrides families

Table S1. Comparison of three families of binary hydrides with high $T_c$.

<table>
<tr><th>hydrogenic motifs</th><th>Hydrides<br>(atomic radius/ electronegativity)<br>Electronic configuration</th><th>Symmetry</th><th>$T_c$ ,K (P, GPa)</th></tr>
<tr><td rowspan="2">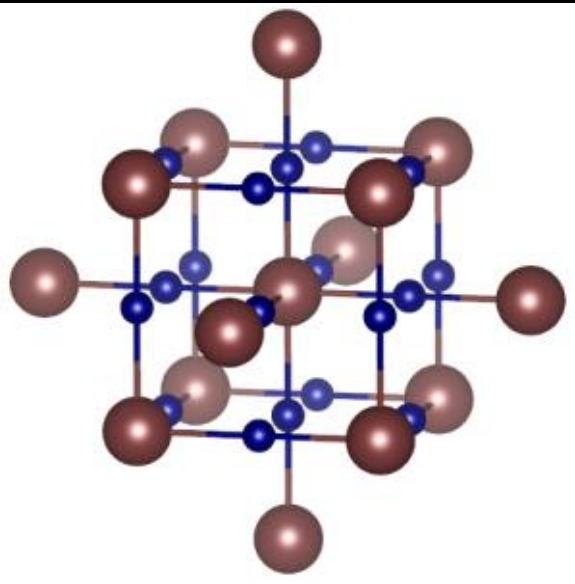<br>H-S(Se) covalent bond</td><td>$H_3S$ (1.84 Å /2.58)<br>[Ne] $3s^2 3p^4$</td><td rowspan="2">$Im\bar{3}m$</td><td>204[a] (200)[32]<br>203[b] (155)[33]</td></tr>
<tr><td>$H_3Se$ (1.98 Å /2.55)<br>[Ne] $3d^{10} 4s^2 4p^4$</td><td>110[a] (200)[34]</td></tr>
<tr><td rowspan="3">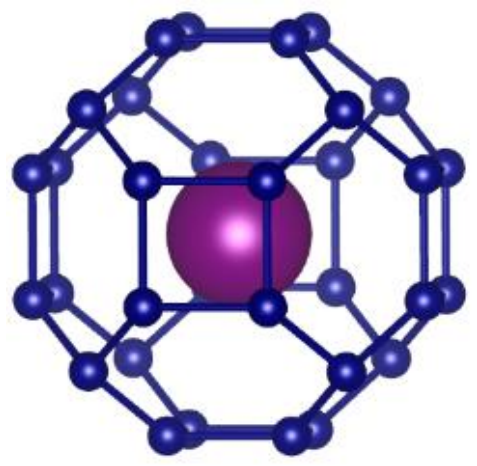<br>$H_{32}$ cage</td><td>$LaH_{10}$ (1.88 Å /1.10)<br>[Xe] $5d^1 6s^2$</td><td rowspan="3">$Fm\bar{3}m$</td><td>288[a] (200)[35]<br>250-260[b]<br>(170-200)[36-37]</td></tr>
<tr><td>$YH_{10}$ (1.81 Å /1.22)<br>[Kr] $4d^1 5s^2$</td><td>303[a] (400)[35]</td></tr>
<tr><td>$ThH_{10}$ (1.80 Å /1.30)<br>[Rn] $6d^2 7s^2$</td><td>241[a] (100)[38]<br>161[b] (174)[39]</td></tr>
<tr><td rowspan="4">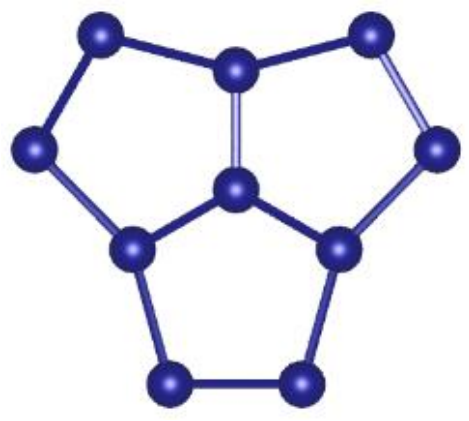<br>$H_{10}$ penta-graphene[c]</td><td>$HfH_{10}$ (1.56 Å /1.30)<br>[Xe] $4f^{14} 5d^2 6s^2$</td><td rowspan="4">$P6_3/mmc$</td><td>234 (250)</td></tr>
<tr><td>$ZrH_{10}$ (1.60 Å /1.33)<br>[Kr] $4d^2 5s^2$</td><td>220 (250)</td></tr>
<tr><td>$ScH_{10}$ (1.61 Å /1.36)<br>[Ar] $3d^1 4s^2$</td><td>158 (250)</td></tr>
<tr><td>$LuH_{10}$ (1.73 Å /1.27)<br>[Xe] $4f^{14} 5d^1 6s^2$</td><td>152 (200)</td></tr>
</table>

[a]$T_c$ values were calculated with $\mu^* = 0.1$.
[b]$T_c$ was measured experimentally.
[c]$T_c$ values were calculated with G-K equation at $\mu^* = 0.1$.

# Thermodynamic stability of Hf-H system

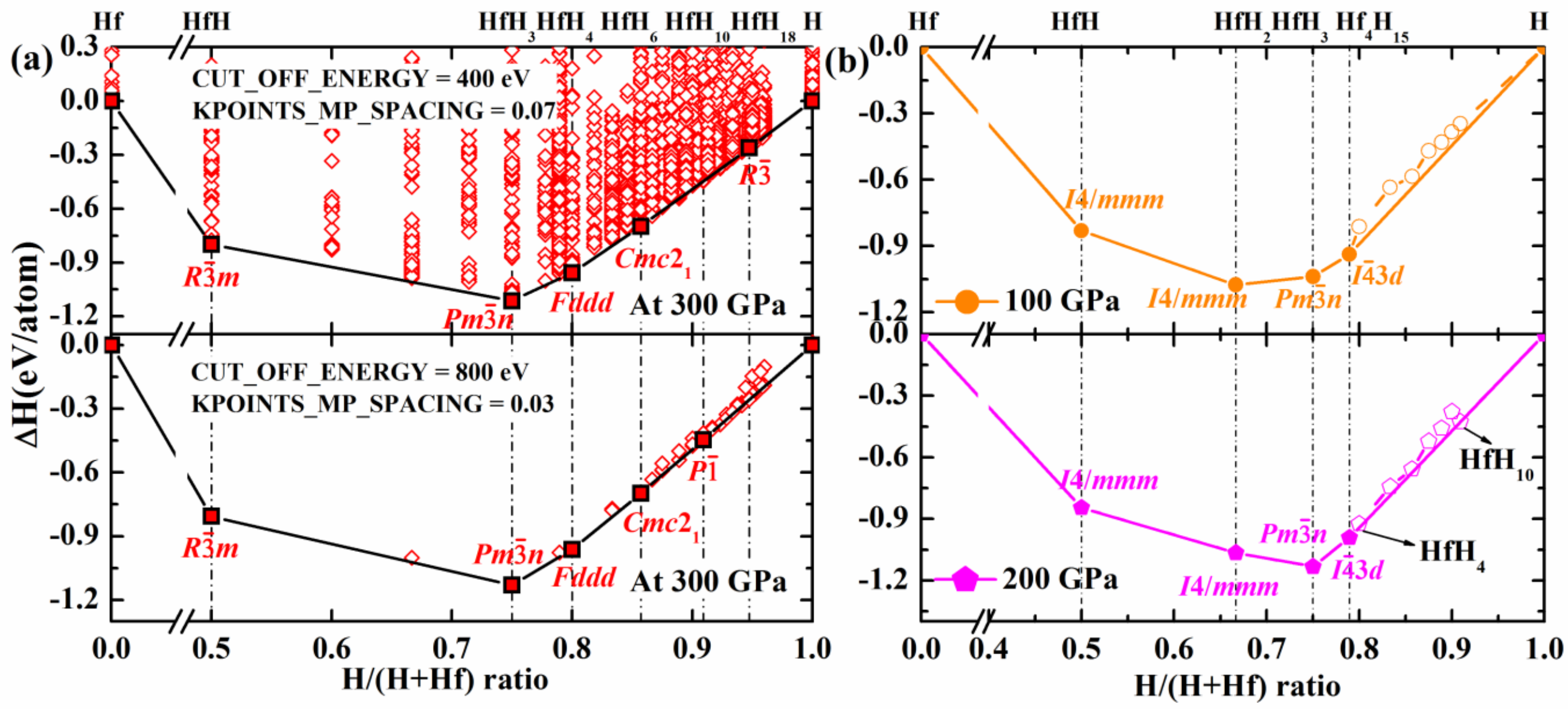

Fig. S1 (a) Predicted formation enthalpies of Hf-H system barring ZPEs with general (upper panel) and high accuracy (nether panel) at 300 GPa. (b) Formation enthalpies of $HfH_x$ (x=1-10) not inclusion ZPE with high accuracy at 100 and 200 GPa. Here, the compounds on the convex hull are thermodynamically stable and, in principle, experimentally synthesizable under corresponding pressures. As can be seen, HfH is predicted to be stable with $I4/mmm$ symmetry from 100 GPa to 200 GPa, then transforms into $R\bar{3}m$ at 300 GPa. $HfH_2$ adopts the $I4/mmm$ structure between 100 GPa and 200 GPa, and then breaks down with respect to a mixture of Hf and $HfH_3$ at 300 GPa. $HfH_3$ is stabilized in a $Pm\bar{3}n$ structure at 100 GPa, and remains as such up to at least 300 GPa. $Hf_4H_{15}$ is predicted to be stable below 200 GPa with space group of $I\bar{4}3d$. $Fddd$-$HfH_4$, $Cmc2_1$-$HfH_6$ and $P\bar{1}$-$HfH_{10}$ are only stable at 300 GPa.

Table S2. Calculated enthalpies and ZPE values of $P\bar{1}$ and $P6_3/mmc$ of $HfH_{10}$ at different pressures. Numbers between parentheses represent the atoms in the supercells using different softwares (VASP/CASTEP).

| Phase ($HfH_{10}$) | P (GPa) | Enthalpy (eV/f.u.) | | ZPE (eV/f.u.) | | Enthalpy + ZPE (eV/f.u.) | |
|---|---|---|---|---|---|---|---|
| | | VASP | CASTEP | VASP | CASTEP | VASP | CASTEP |
| $P\bar{1}$ (88/88) | 200 | 1.3108 | -7977.4660 | 2.9915 | 3.0468 | 4.3023 | -7974.4192 |
| | 250 | 9.8058 | -7968.9748 | 3.1177 | 3.1697 | 12.9235 | -7965.8051 |
| | 300 | 17.7743 | -7961.0109 | 3.2245 | 3.2779 | 20.9988 | -7957.7330 |
| $P6_3/mmc$ (66/44) | 200 | 1.9724 | -7976.7688 | 2.5059 | 2.5712 | 4.4783 | -7974.1976 |
| | 250 | 10.2612 | -7968.4730 | 2.7255 | 2.7245 | 12.9867 | -7965.7485 |
| | 300 | 18.0640 | -7960.6644 | 2.9132 | 2.8890 | 20.9772 | -7957.7754 |
| ΔE [$P\bar{1}$-$P6_3/mmc$] | 200 | -0.6616 | -0.6972 | 0.4856 | 0.4756 | -0.1760 | -0.2216 |
| | 250 | -0.4554 | -0.5018 | 0.3922 | 0.4452 | -0.0632 | -0.0566 |
| | 300 | -0.2897 | -0.3465 | 0.3113 | 0.3889 | 0.0216 | 0.0424 |

## Structures and bond analysis of Hf-H system

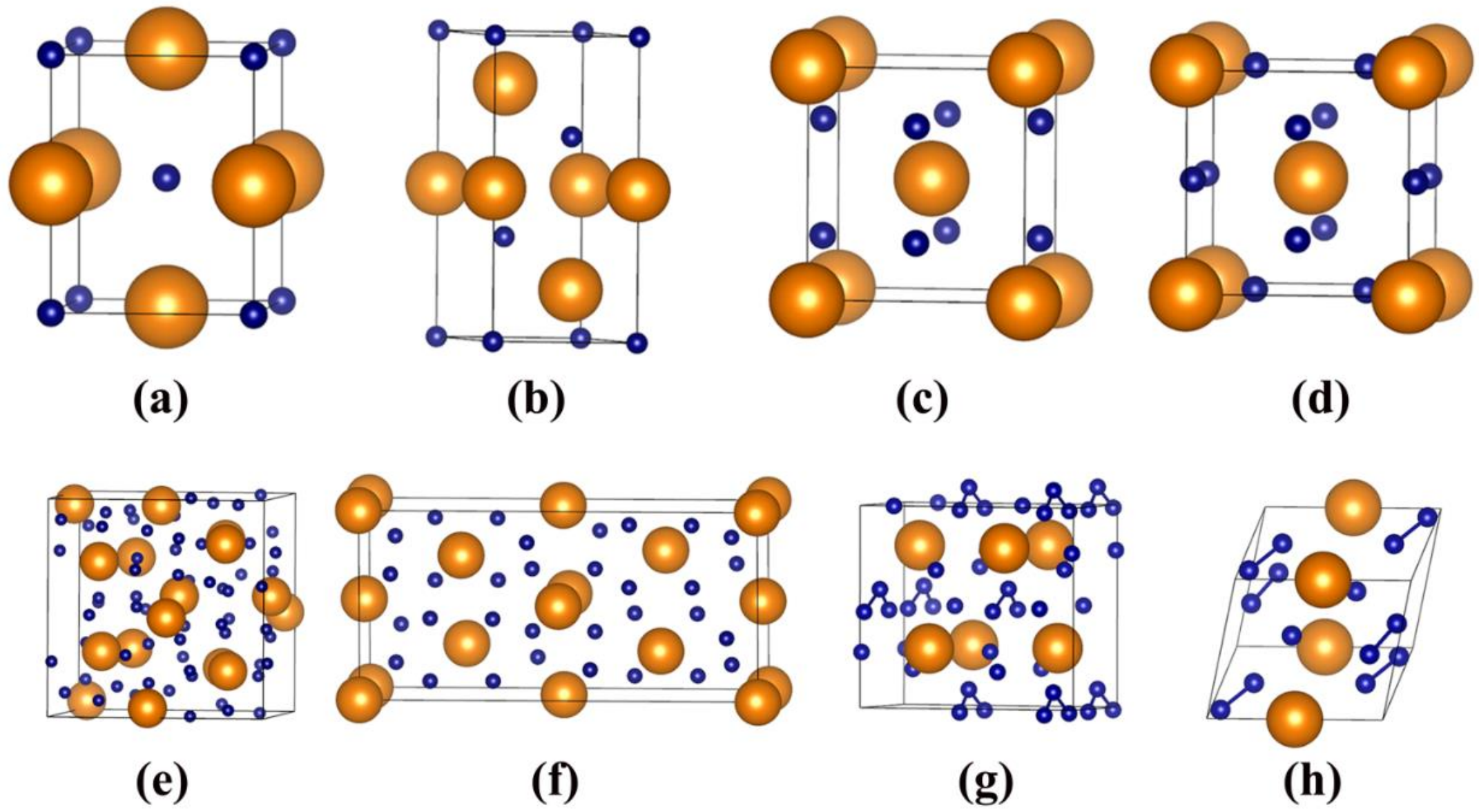

Fig. S2. Structures of stable Hf-H compounds under high pressures. (a) $I4/mmm$ in HfH, (b) $R\bar{3}m$ in HfH, (c) $I4/mmm$ in $HfH_2$, (d) $Pm\bar{3}n$ in $HfH_3$, (e) $R\bar{3}c$ in $HfH_3$, (f) $I\bar{4}3d$ in $Hf_4H_{15}$, (g) $Fddd$ in $HfH_4$ (h) $Cmc2_1$ in $HfH_6$, (i) $P\bar{1}$ in $HfH_{10}$. Large golden spheres denote Hf atoms, while the small blue spheres represent H atoms.

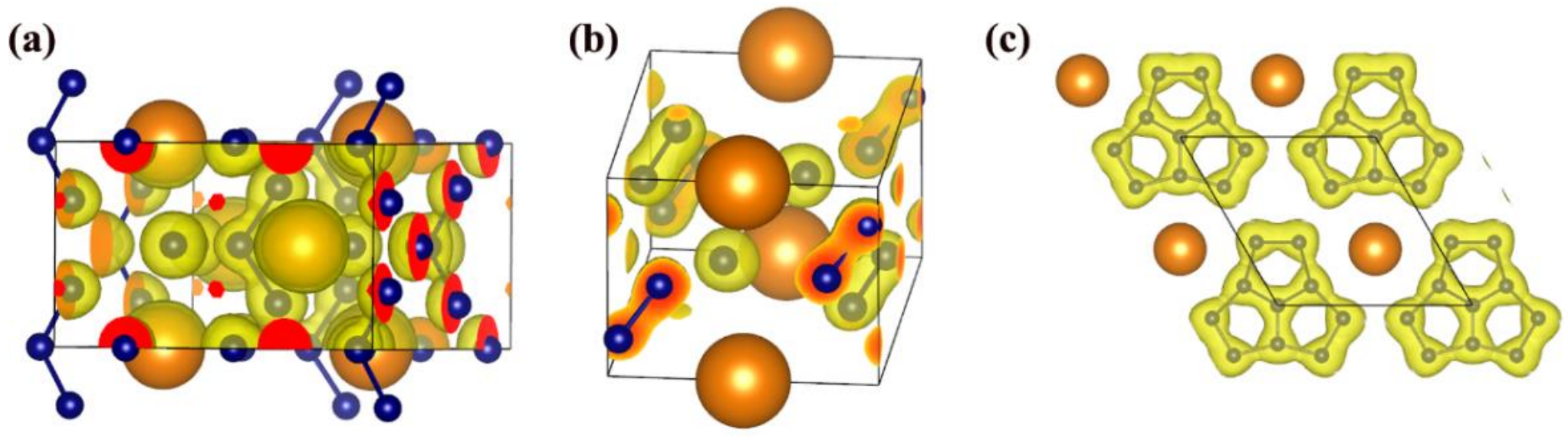

Fig. S3. The calculated ELF of (a) *Cmc*$2_1$-$HfH_6$ with isosurface value of 0.7 at 300 GPa, (b) $P\bar{1}$-$HfH_{10}$ with isosurface value of 0.8 at 200 GPa, (c) $P6_3/mmc$-$HfH_{10}$ with isosurface value of 0.6 at 300 GPa. For $HfH_6$, The H-H bond lengths in the $H_3$ unit are approximately 0.97 Å and 1.05 Å at 300 GPa. The electron localization function (ELF) values of the H-H bonds in $HfH_6$ reach 0.7-0.8, indicating strong covalent bonding characteristics. The $P\bar{1}$ structure of $HfH_{10}$ consists of diatomic hydrogen pairs similar to $H_2$ molecules in which the H-H distances are 0.84 and 0.91 Å and ELF values greater than 0.8 at 200 GPa, which also corresponds to strong covalent bonding.

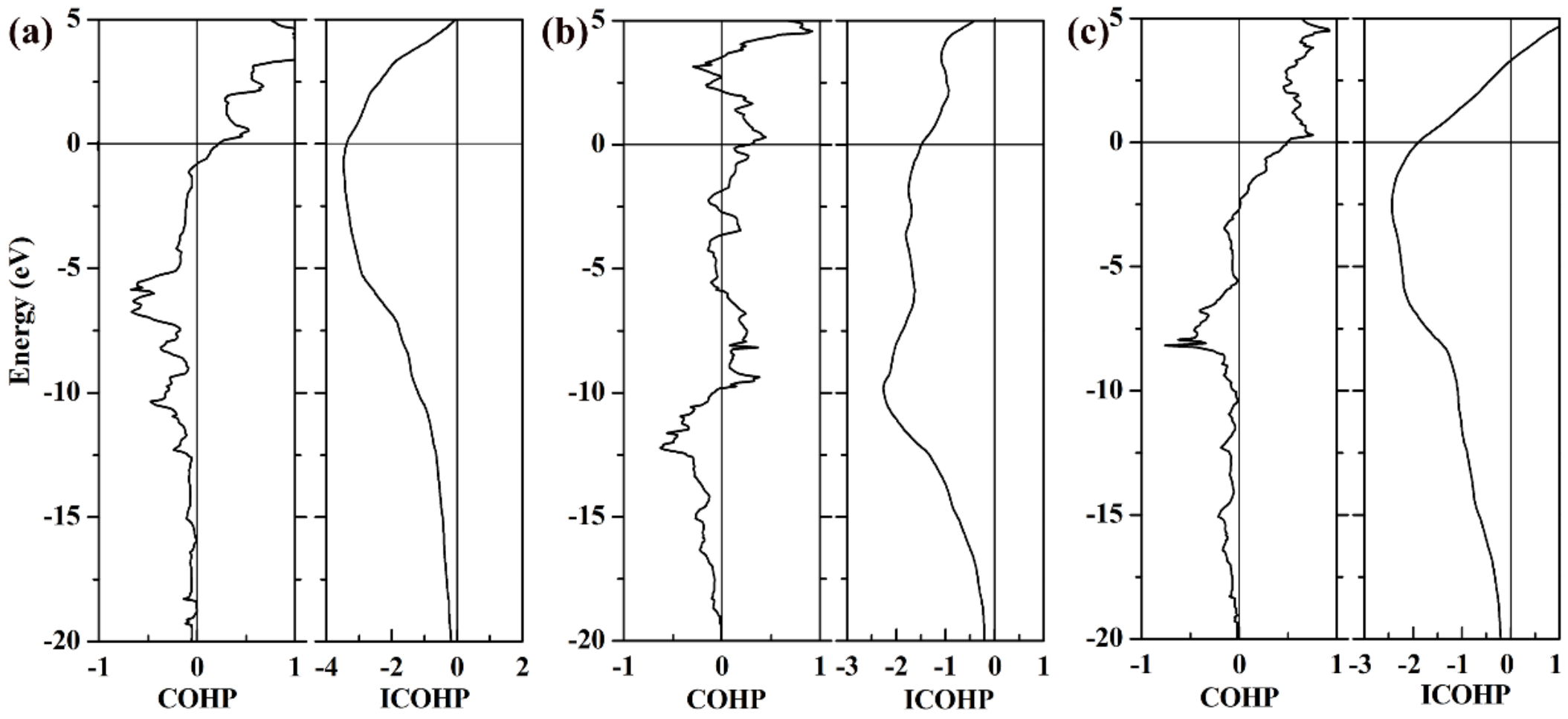

Fig. S4. The calculated Crystalline Orbital Hamiltonian Population (COHP) and Integrated Crystalline Orbital Hamiltonian Population (ICOHP) of $P6_3/mmc$-$HfH_{10}$ with H-H distances of (a) 0.92, (b) 1.02 and (c) 1.07 Å at 300 GPa. The horizontal lines present the Fermi levels. The negative COHP indicates bonding and positive COHP indicates antibonding. And the negative ICOHP values represent the bonding interactions between the H atoms. The fully occupied bonding states and partially occupied antibonding states lends a strong support on the H-H covalent bonding within the planar $H_{10}$ units.

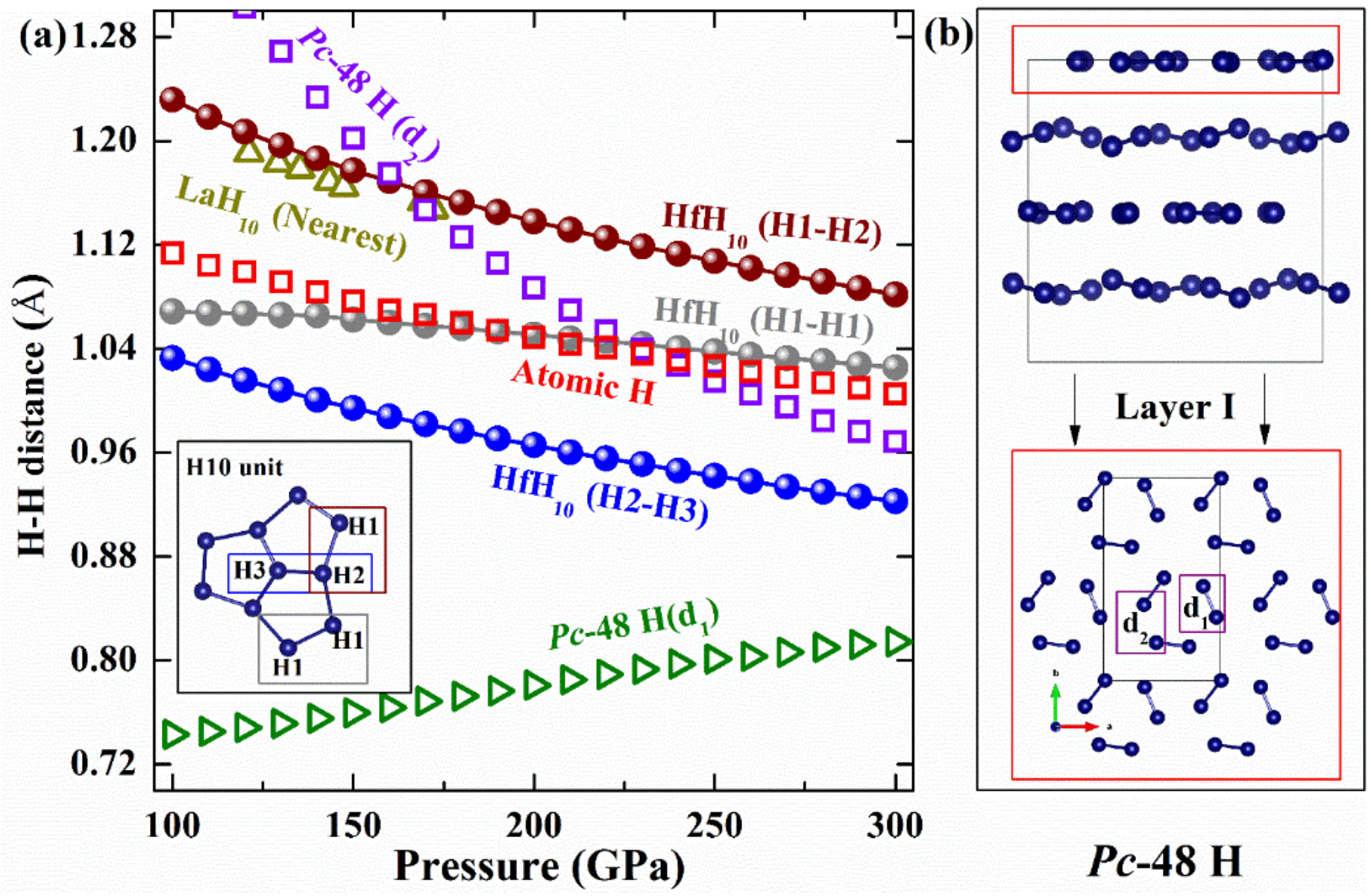

Fig. S5. (a) The calculated H-H distances in $HfH_{10}$ in comparison to $LaH_{10}$, atomic H and *Pc*-48 H in a pressure range of 100-300 GPa. Inset: schematic diagram of $H_{10}$ unit in $HfH_{10}$. (b) Side view of the *Pc*-48 H structure and top view of a weakly bonded graphene-like layer (layer I).

# Thermodynamic stability of Zr-H system

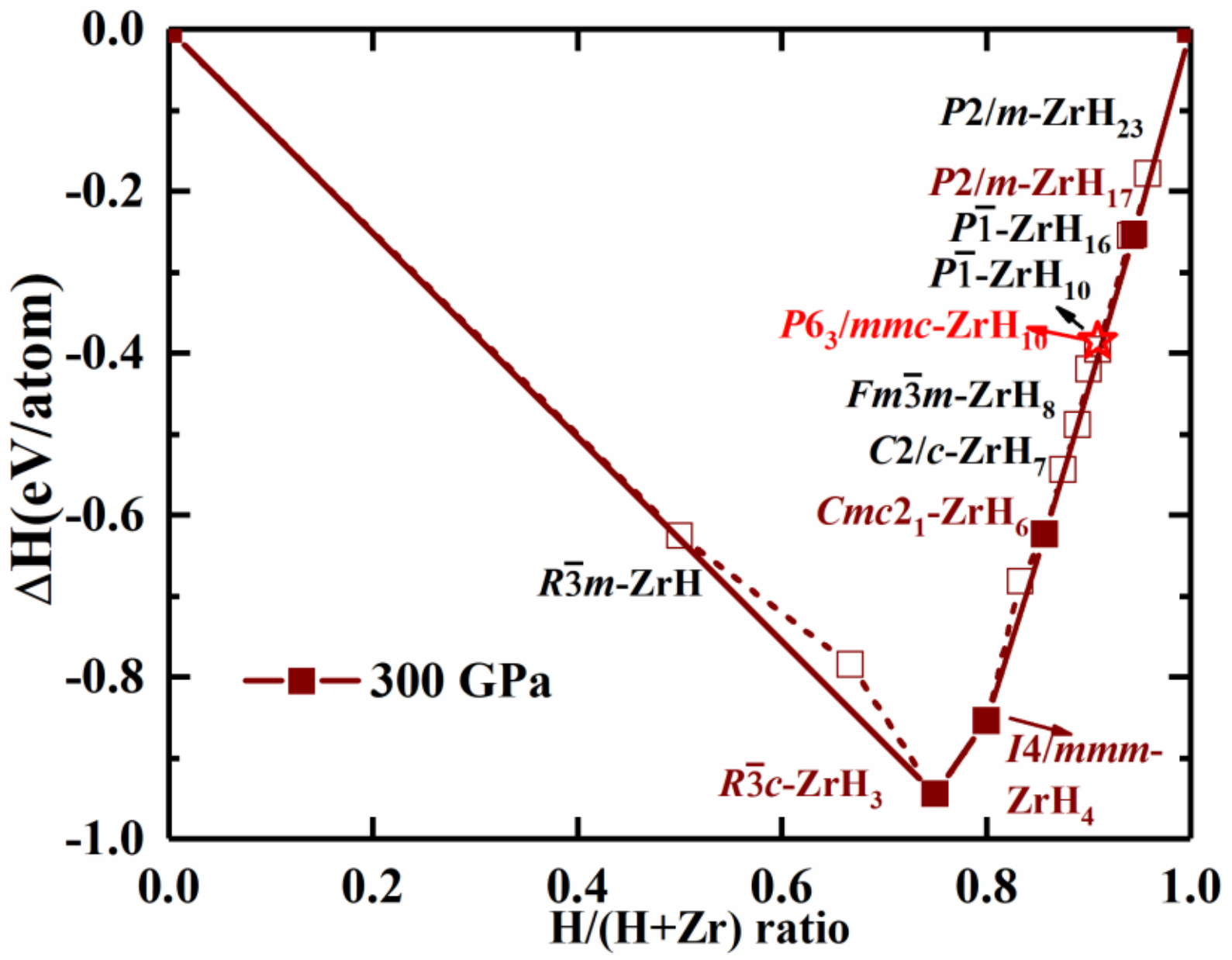

Fig. S6. Predicted formation enthalpies of Zr-H system with decomposition into elemental zirconium and hydrogen at 300 GPa. It is found that $ZrH_3$, $ZrH_4$, $ZrH_6$ and $ZrH_{17}$ are thermodynamically stable, while $ZrH_{10}$ is metastable. For $ZrH_{10}$, *P*$6_3$/*mmc* and *P*$\bar{1}$ phases are competitive phase with 18 meV/atom and 8 meV/atom above the convex hull, respectively.

Table S3. Calculated enthalpies and ZPE values of $ZrH_{10}$ structures. Numbers between parentheses represent the atoms in the supercells using different softwares (VASP/CASTEP).

| Phase ($ZrH_{10}$) | P (GPa) | Enthalpy (eV/f.u.) | | ZPE (eV/f.u.) | | Enthalpy + ZPE (eV/f.u.) | |
|---|---|---|---|---|---|---|---|
| | | VASP | CASTEP | VASP | CASTEP | VASP | CASTEP |
| $P\bar{1}$ (88/44) | 200 | 5.8245 | -1403.6864 | 2.7388 | 2.8256 | 8.5633 | -1400.8608 |
| | 250 | 14.2116 | -1395.3404 | 2.8666 | 2.9478 | 17.0782 | -1392.3926 |
| | 300 | 22.0854 | -1387.5206 | 2.9863 | 3.0713 | 25.0717 | -1384.4493 |
| $P6_3/mmc$ (66/44) | 200 | 5.9177 | -1403.5396 | 2.5975 | 2.6115 | 8.5152 | -1400.9281 |
| | 250 | 14.2794 | -1395.2253 | 2.7879 | 2.8030 | 17.0673 | -1392.4223 |
| | 300 | 22.1523 | -1387.4095 | 2.9397 | 2.9316 | 25.0920 | -1384.4779 |
| ΔE [$P\bar{1}$-$P6_3/mmc$] | 200 | -0.0932 | -0.1468 | 0.1413 | 0.2141 | 0.0481 | 0.0673 |
| | 250 | -0.0678 | -0.1151 | 0.0787 | 0.1448 | 0.0109 | 0.0297 |
| | 300 | -0.0669 | -0.1111 | 0.0466 | 0.1397 | -0.0203 | 0.0286 |

# Superconductive parameters of $MH_{10}$ (M=Hf, Zr, Sc and Lu)

Table S4. The calculated EPC parameter $\lambda$, logarithmic average phonon frequency $\omega_{log}$ (K), electronic density of states at Fermi level $N(\varepsilon_f)$ (states/spin/Ry/f.u.) and superconducting transition temperatures $T_c$ (K) with $\mu^*$=0.1-0.13 at corresponding pressures $P$ (GPa).

| Structure | $P$ | $\lambda$ | $\omega_{log}$ | $N(\varepsilon_f)$ | $T_c$ (K)[a] | A-D $T_c$ (K)[b] | LE $T_c$ (K) | G-K $T_c$ (K) |
|---|---|---|---|---|---|---|---|---|
| $P6_3/mmc$-$HfH_{10}$ | 250 | 2.77 | 677.3 | 6.5 | 112-118 | 152-167 | 226-239 | 213-234 |
| | 300 | 2.16 | 861.5 | 6.2 | 122-130 | 151-166 | 214-228 | 197-220 |
| $P6_3/mmc$-$ZrH_{10}$ | 250 | 1.77 | 1068.8 | 6.4 | 130-141 | 151-167 | 185-198 | 199-220 |
| | 300 | 1.59 | 1162.6 | 6.1 | 128-140 | 145-162 | 177-191 | 194-218 |
| $P6_3/mmc$-$ScH_{10}$ | 250 | 1.16 | 1211.3 | 3.6 | 92-104 | 99-114 | 112-124 | 134-158 |
| $P6_3/mmc$-$LuH_{10}$ | 200 | 1.36 | 1024.2 | 3.4 | 95-105 | 105-118 | 123-134 | 134-152 |
| $Fm\bar{3}m$-$LaH_{10}$ | 200 | 4.36 | 655.6 | 5.0 | 130-135 | 214-234 | 264-276 | 252-271 |
| | 300 | 1.94 | 1380.7 | 5.4 | 181-195 | 211-231 | 250-264 | 210-232 |
| $Fm\bar{3}m$-$YH_{10}$ | 300 | 2.02 | 1444.0 | 4.6 | 196-209 | 231-253 | 277-292 | 235-259 |
| $Im\bar{3}m$-$H_3S$ | 200 | 1.88 | 1266.0 | 3.1 | 162-175 | 186-204 | 212-223 | 206-229 |

[a] $T_c$ was estimated using Allen-Dynes modified McMillian equation with $f_1 f_2 = 1$ (Eq. S11).

[b] $T_c$ was estimated using Allen-Dynes modified McMillian equation with $f_1 f_2 \neq 1$ (Eq. S12).

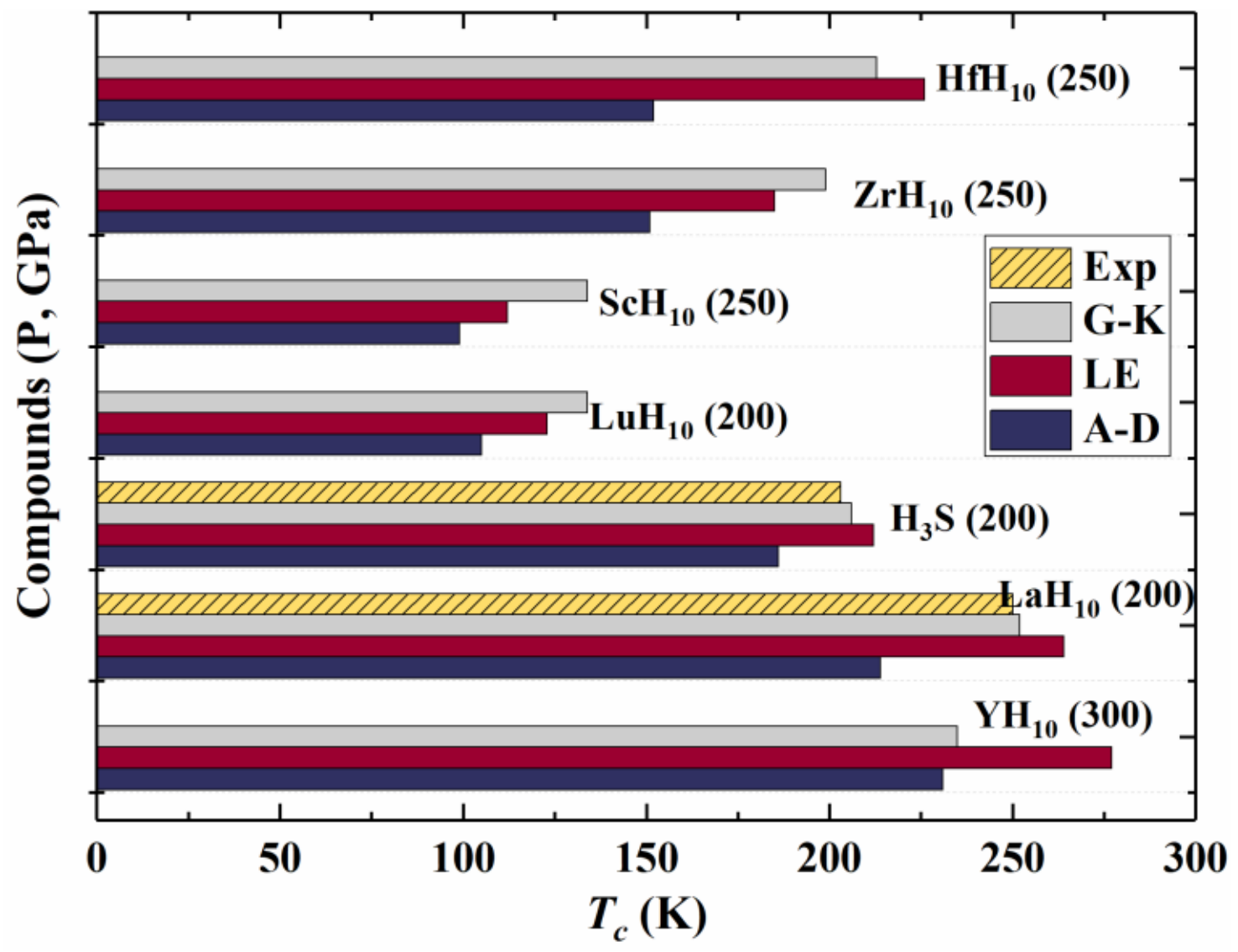

Fig. S7. Comparison of $T_c$ ($\mu^*$=0.13) values obtained by different equations for different hydrogen-rich materials.

Table S5. Calculated isotope effect and superconductive gap of $MH_{10}$ (M=Hf, Zr, Sc, Lu).

| Parameter | $HfH_{10}$ | $ZrH_{10}$ | $ScH_{10}$ | $LuH_{10}$ |
|---|---|---|---|---|
| | 250 GPa | 250 GPa | 250 GPa | 200 GPa |
| α | 0.42-0.43 | 0.38-0.39 | 0.37-0.38 | 0.44-0.45 |
| $T_c^D$, K | 159-174 | 153-168 | 104-121 | 99-111 |
| Δ(0), meV | 39.9-45.2 | 35.3-40.0 | 22.1-26.6 | 22.4-25.8 |

# Projected DOS and superconductive parameters of the other hafnium hydrides

For $HfH_x$ compounds with $x$ = 1 to 4, the hydrogen contribution to the DOS is relatively low at the Fermi level, which results in moderate electron-phonon coupling (EPC). For $HfH_6$ and $P\bar{1}$-$HfH_{10}$, the total DOS at the Fermi level is low, due to orbital splitting during the formation of diatomic/triatomic hydrogen units in the structure. The estimated $T_c$'s for $HfH_{1\text{-}3}$, as expected, are fairly low with values of < 20 K due to the weak electron-phonon interaction and low logarithmic average frequency ($\omega_{log}$). As the hydrogen content increases, the EPC is strengthened and $\omega_{log}$ enhanced. Electron-phonon coupling calculations for $HfH_4$, $HfH_6$ and $P\bar{1}$-$HfH_{10}$ yield $\lambda$ of 0.88 (200 GPa), 0.84 (300 GPa) and 0.72 (200 GPa), respectively. The calculated $T_c$'s are increased to 41.6-50.1, 45.2-55.0 and 28.9-37.4

K using a typical Coulomb potential of $\mu^*$ = 0.1-0.13. Although $HfH_{10}$ has the highest hydrogen content, the existence of $H_2$ units in the $P\bar{1}$ structure reduces the electronic density of states at the Fermi energy, thereby limiting its superconductivity

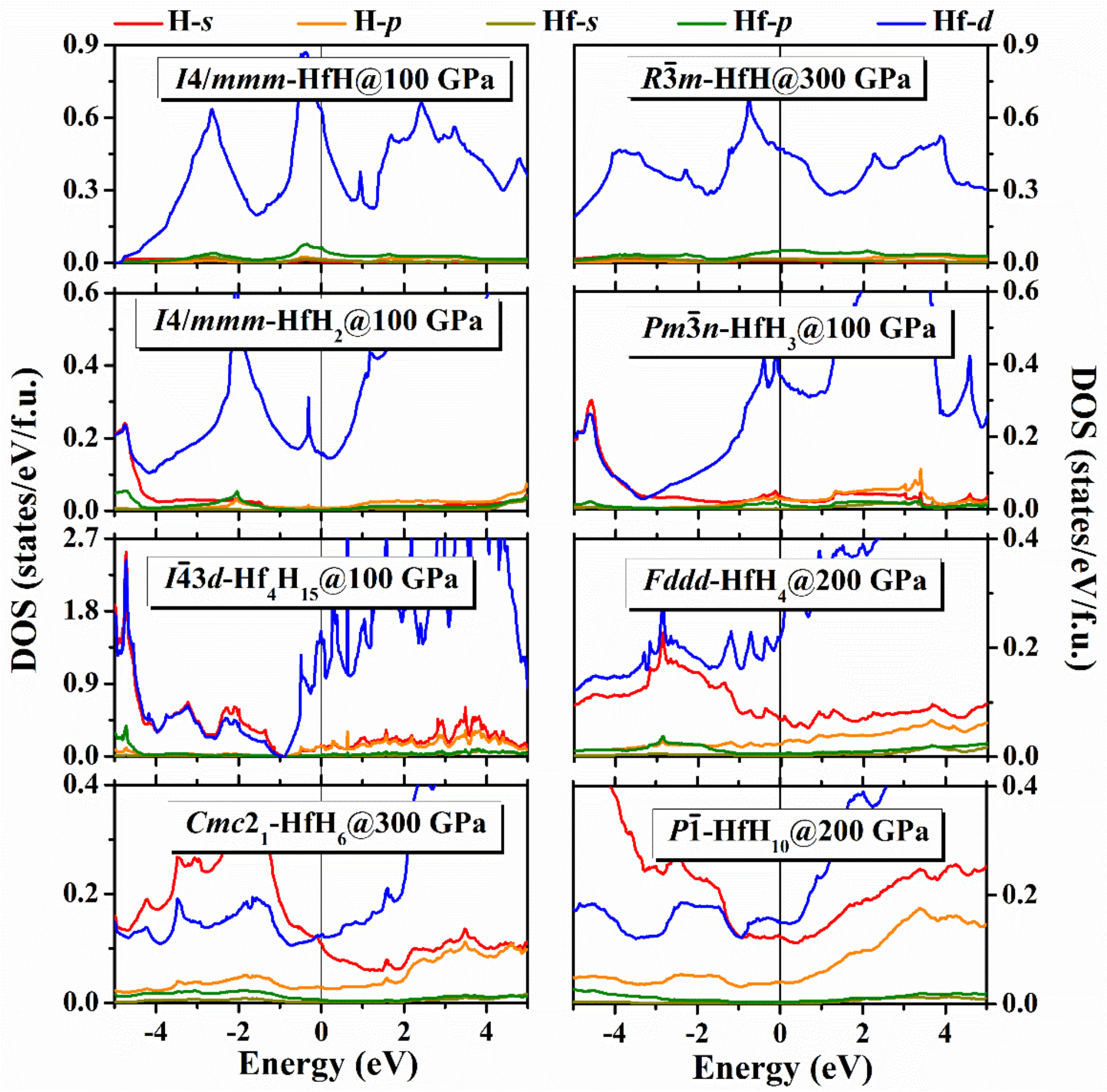

Fig. S8. Calculated projected density of states (DOS) for hafnium hydrides under pressures. All compounds were found to be metallic with the absence of a band gap.

Table S6. The calculated EPC parameter λ, logarithmic average phonon frequency $\omega_{log}$, electronic density of states at Fermi level $N(\varepsilon_f)$ (states/spin/Ry/f.u.) and superconducting transition temperatures $T_c$ (K) with $\mu^*$=0.1-0.13 at corresponding pressures *P* (GPa).

| Structure | P | λ | $\omega_{log}$ (K) | N(εf) | $T_c$ (K) |
|---|---|---|---|---|---|
| *I*4/*mmm*-HfH | 100 | 0.68 | 270.8 | 6.5 | 6.8-8.9 |
| $R\bar{3}m$-HfH | 300 | 1.24 | 157.6 | 5.2 | 13.0-14.6 |
| *I*4/*mmm*-$HfH_2$ | 100 | 0.15 | 382.8 | 2.1 | 0 |
| $Pm\bar{3}n$-$HfH_3$ | 100 | 0.62 | 582.6 | 4.3 | 10.6-14.7 |
| $I\bar{4}3d$-$Hf_4H_{15}$ | 100 | 0.37 | 791.1 | 12.1 | 0.8-2.1 |
| *Fddd*-$HfH_4$ | 200 | 0.88 | 892.3 | 3.3 | 41.4-50.1 |
| *Cmc*$2_1$-$HfH_6$ | 300 | 0.84 | 1057.2 | 2.5 | 45.2-55.0 |
| $P\bar{1}$-$HfH_{10}$ | 200 | 0.72 | 1013.9 | 3.3 | 28.9-37.4 |
| | 300 | 0.66 | 1207.4 | 2.9 | 26.8-36.1 |

# $MgH_{10}$ and $ThH_{10}$

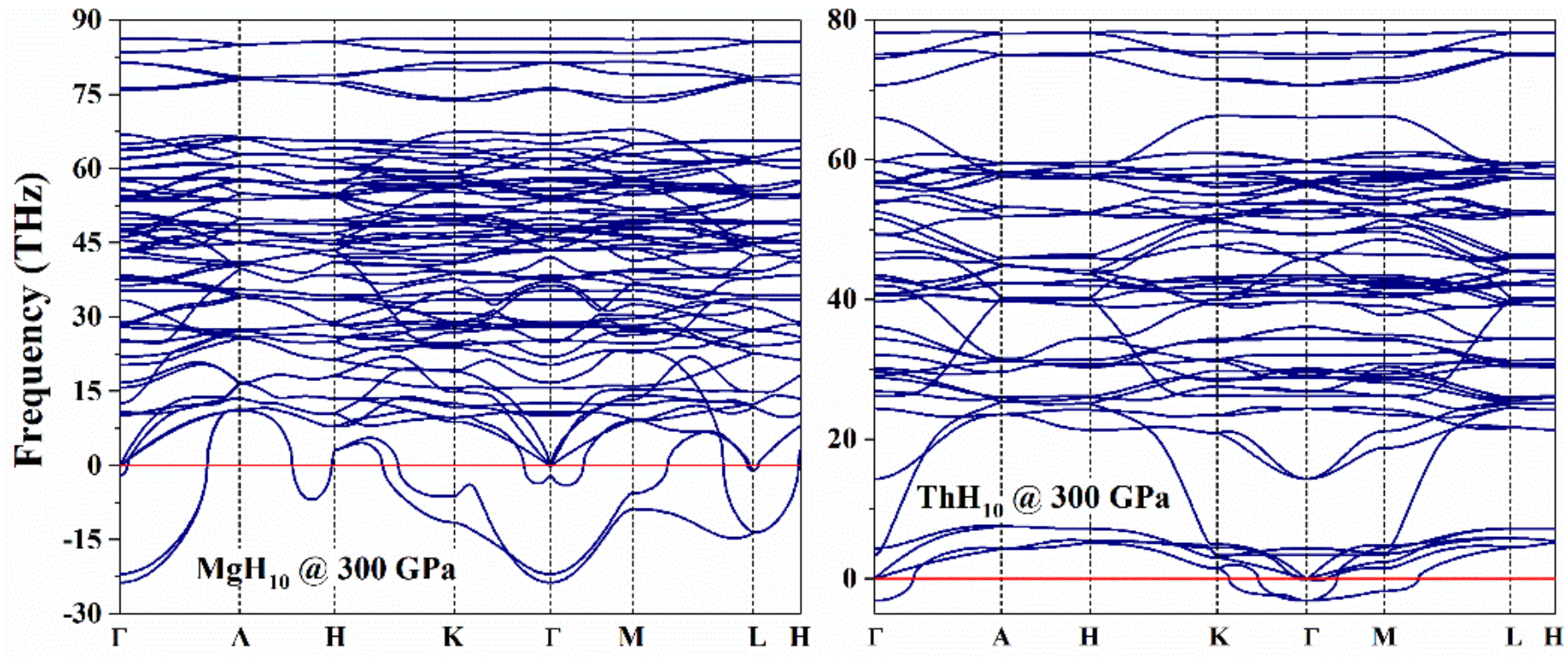

Fig. S9. Phonon spectrum of $P6_3/mmc$-$MgH_{10}$ and $P6_3/mmc$-$ThH_{10}$ at 300 GPa. There are imaginary phonon frequencies for the two structures, indicating their dynamic instabilities

# $ScH_{10}$

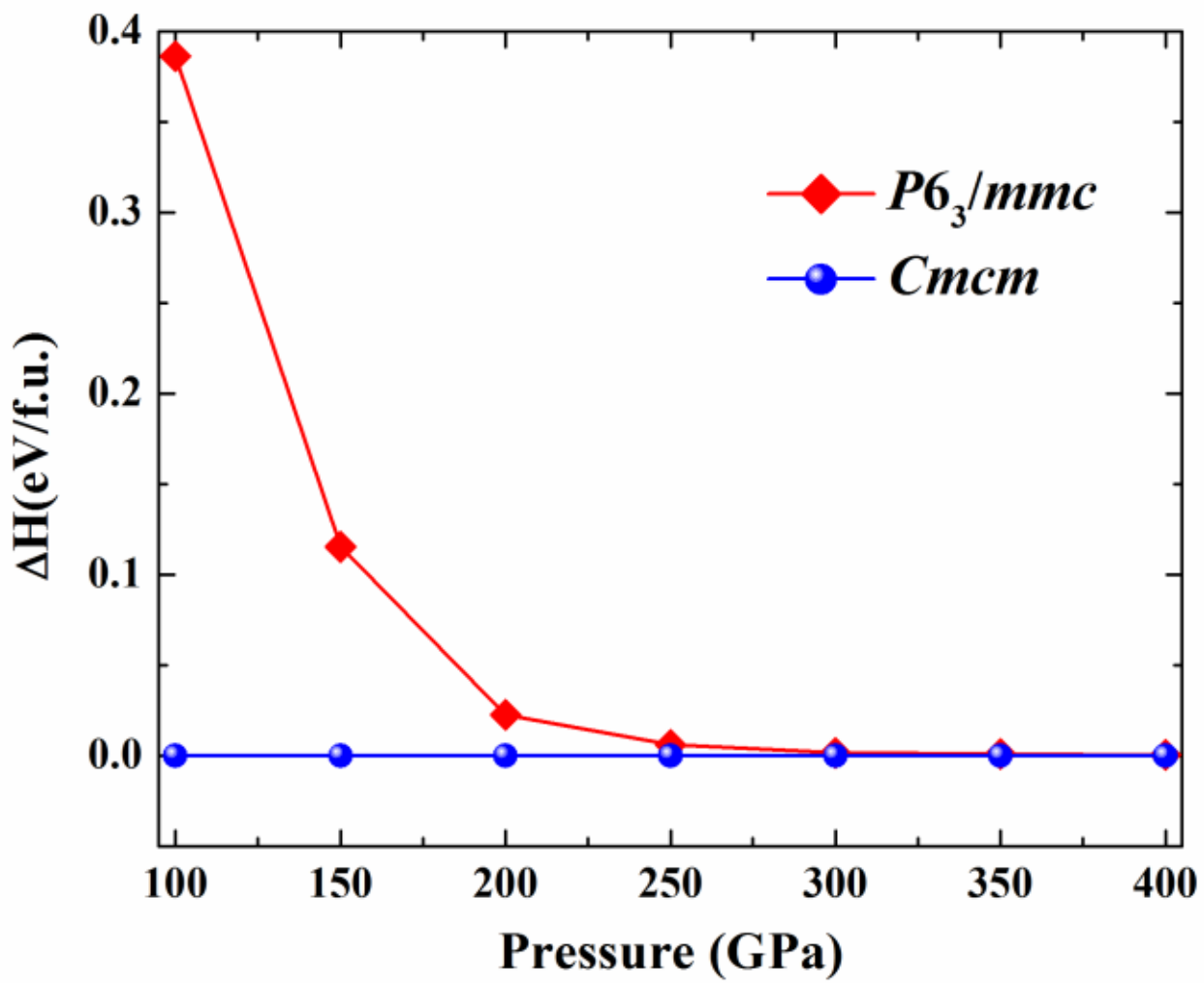

Fig. S10. Calculated enthalpy of $P6_3/mmc$ structure relative to *Cmcm* in $ScH_{10}$.

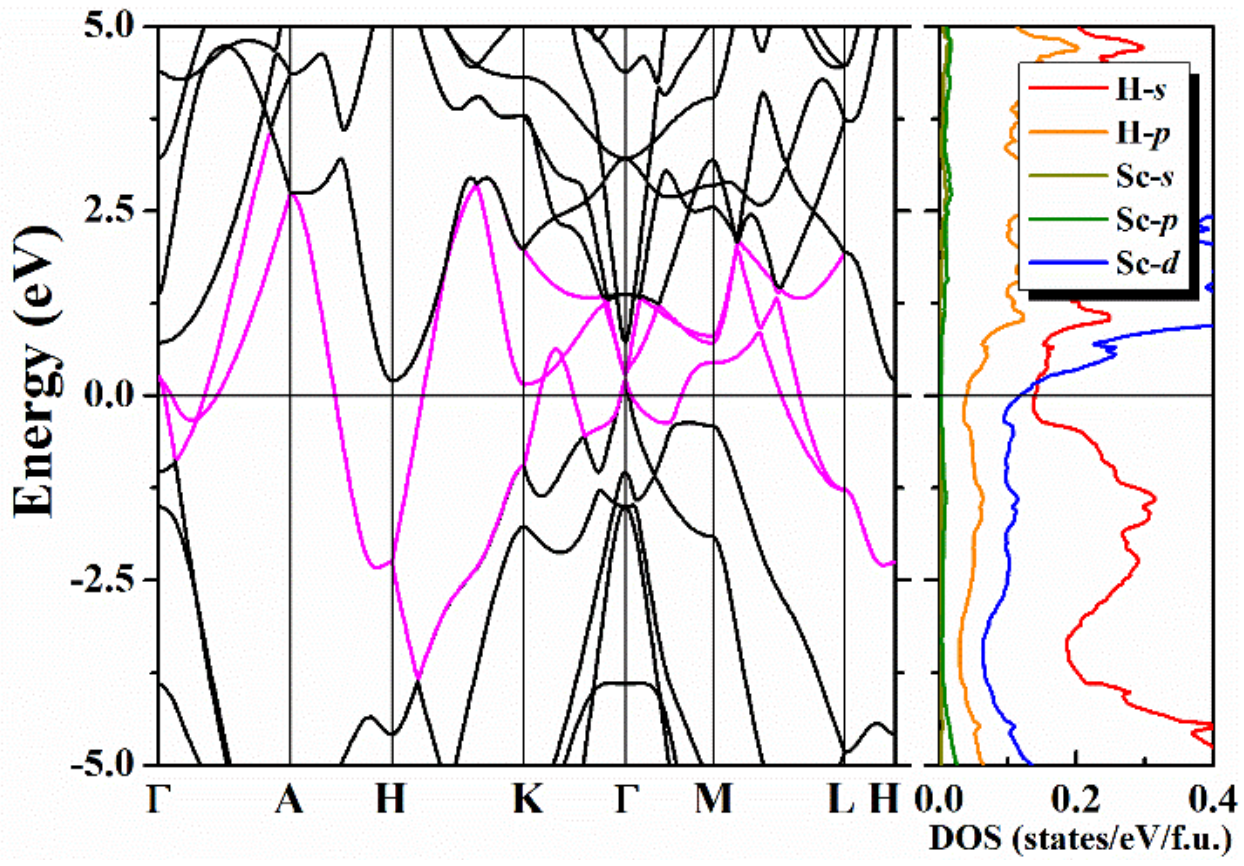

Fig. S11. Calculated electronic band structures and projected DOS of $P6_3/mmc$-$ScH_{10}$ at 250 GPa.

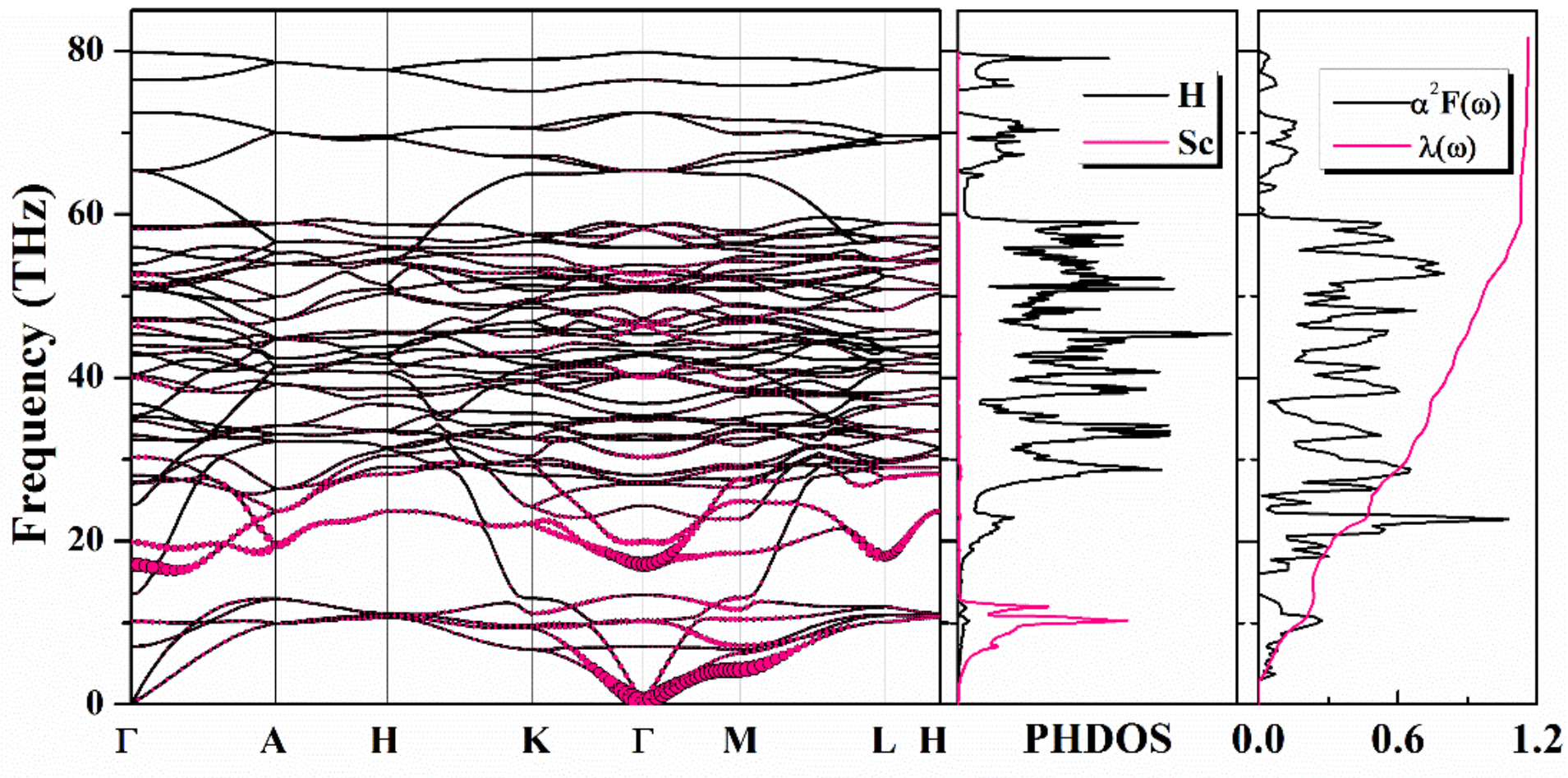

Fig. S12. Phonon dispersion curves, phonon density of states and Eliashberg spectral function $\alpha^2F(\omega)$ together with the electron-phonon integral $\lambda(\omega)$ of $P6_3/mmc$-$ScH_{10}$ at 250 GPa. The absence of imaginary frequency confirms its dynamic stability.

## $LuH_{10}$

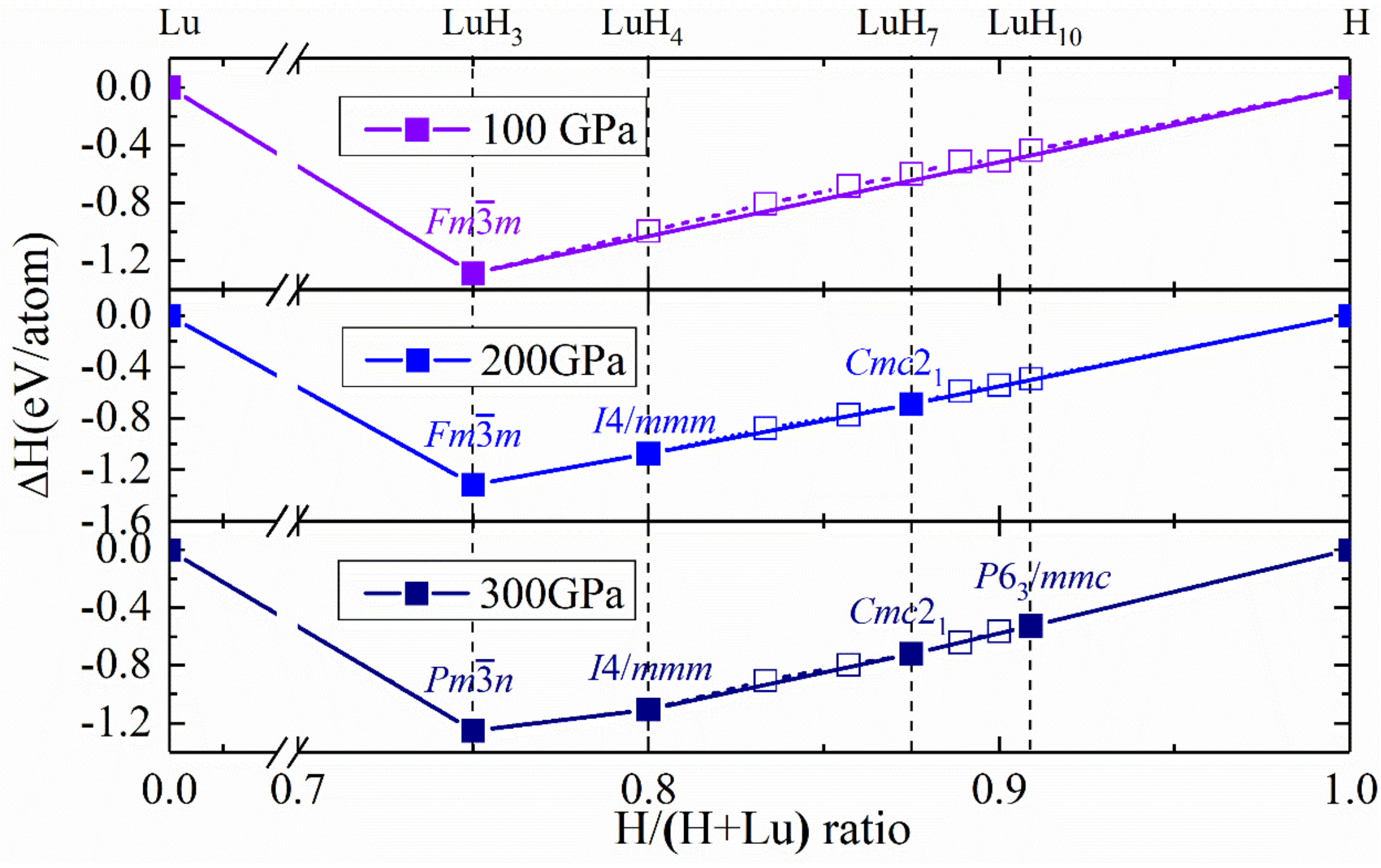

Fig. S13. Predicted formation enthalpies of $LuH_n$ (n=3-10) not inclusion ZPE with respect to decomposition into Lu and $H_2$ under pressure. $LuH_3$ is predicted to be stable with *Fm*$\bar{3}$*m* symmetry from 100 GPa to 200 GPa, then transforms into *Pm*$\bar{3}$*n* at 300 GPa. $LuH_4$ is stabilized in a *I*4/*mmm* structure at 200 GPa, and remains such so up to at least 300 GPa. $LuH_7$ adopts the *Cmc*$2_1$ structure in the pressure range of 200-300 GPa. *P*$6_3$/*mmc*-$LuH_{10}$ is only stable at 300 GPa.

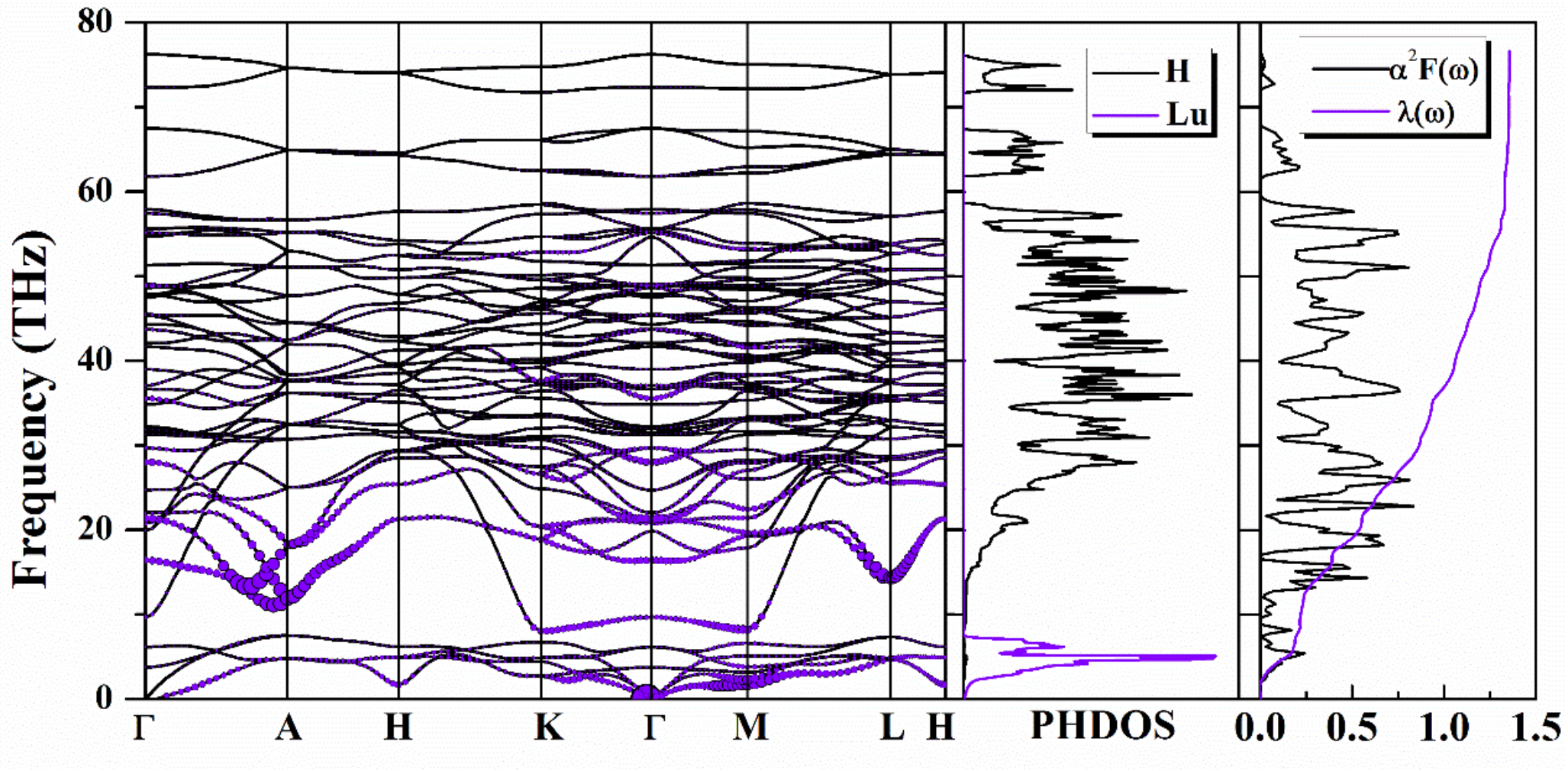

Fig. S14. Phonon dispersion curves, phonon density of states and Eliashberg spectral function $\alpha^2F(\omega)$ together with the electron-phonon integral $\lambda(\omega)$ of *P*$6_3$/*mmc*-$LuH_{10}$ at 200 GPa. The absence of imaginary frequency confirms its dynamic stability.

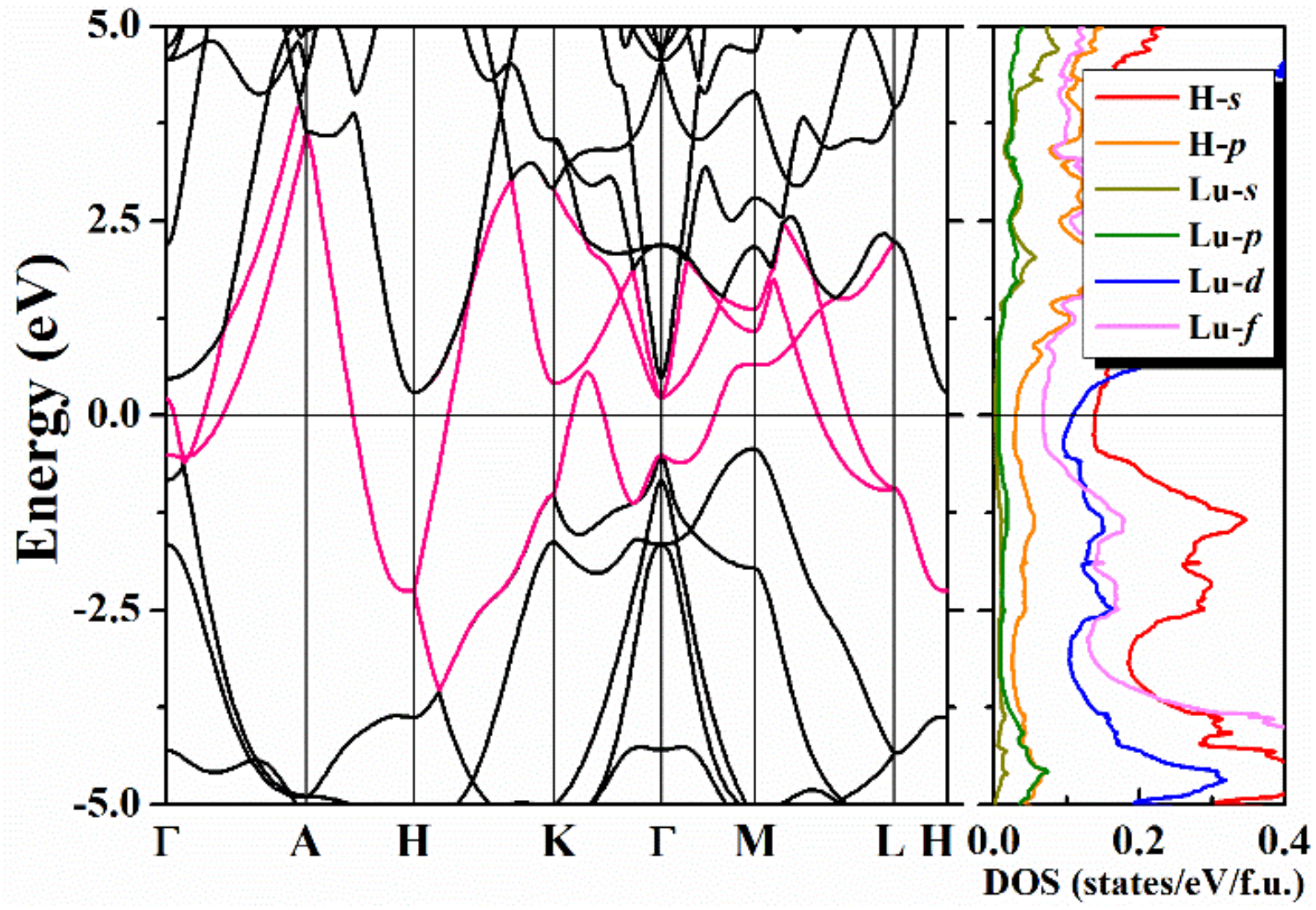

Fig. S15. Calculated electronic band structures and projected DOS of $P6_3/mmc$-$LuH_{10}$ at 200 GPa.

# Dynamic stability

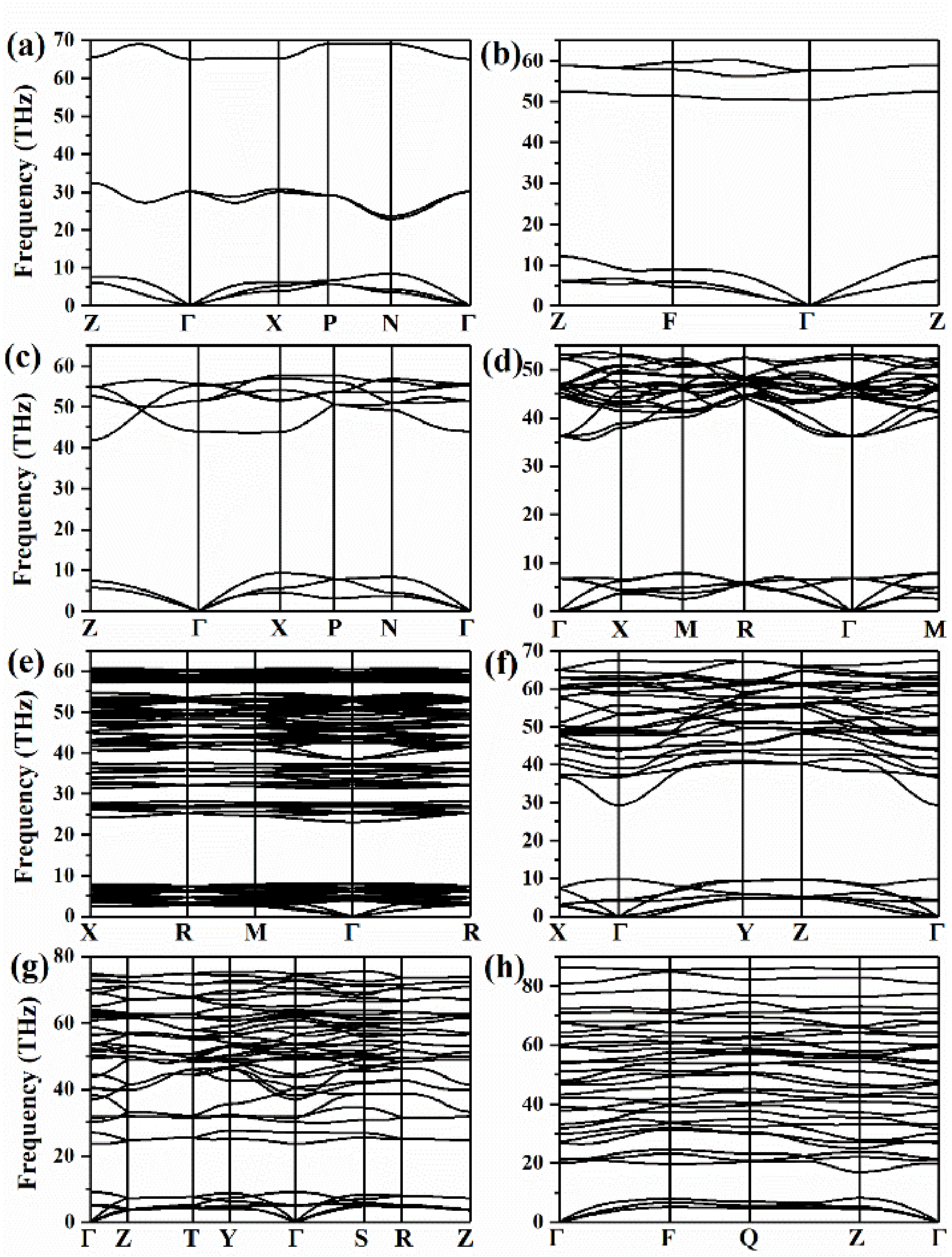

Fig. S16. The calculated phonon dispersion curves of (a) $I4/mmm$-HfH at 100 GPa, (b) $R\bar{3}m$-HfH, (c) $I4/mmm$-$HfH_2$ at 100 GPa, (d) $Pm\bar{3}n$-$HfH_3$ at 100 GPa, (e) $I\bar{4}3d$-$Hf_4H_{15}$ at 100 GPa, (f) $Fddd$-$HfH_4$ at 300 GPa, (g) $Cmc2_1$-$HfH_6$ at 300 GPa, (h) $P\bar{1}$-$HfH_{10}$ at 300 GPa. The absence of imaginary frequency of Hf hydrides confirms their dynamic stabilities.

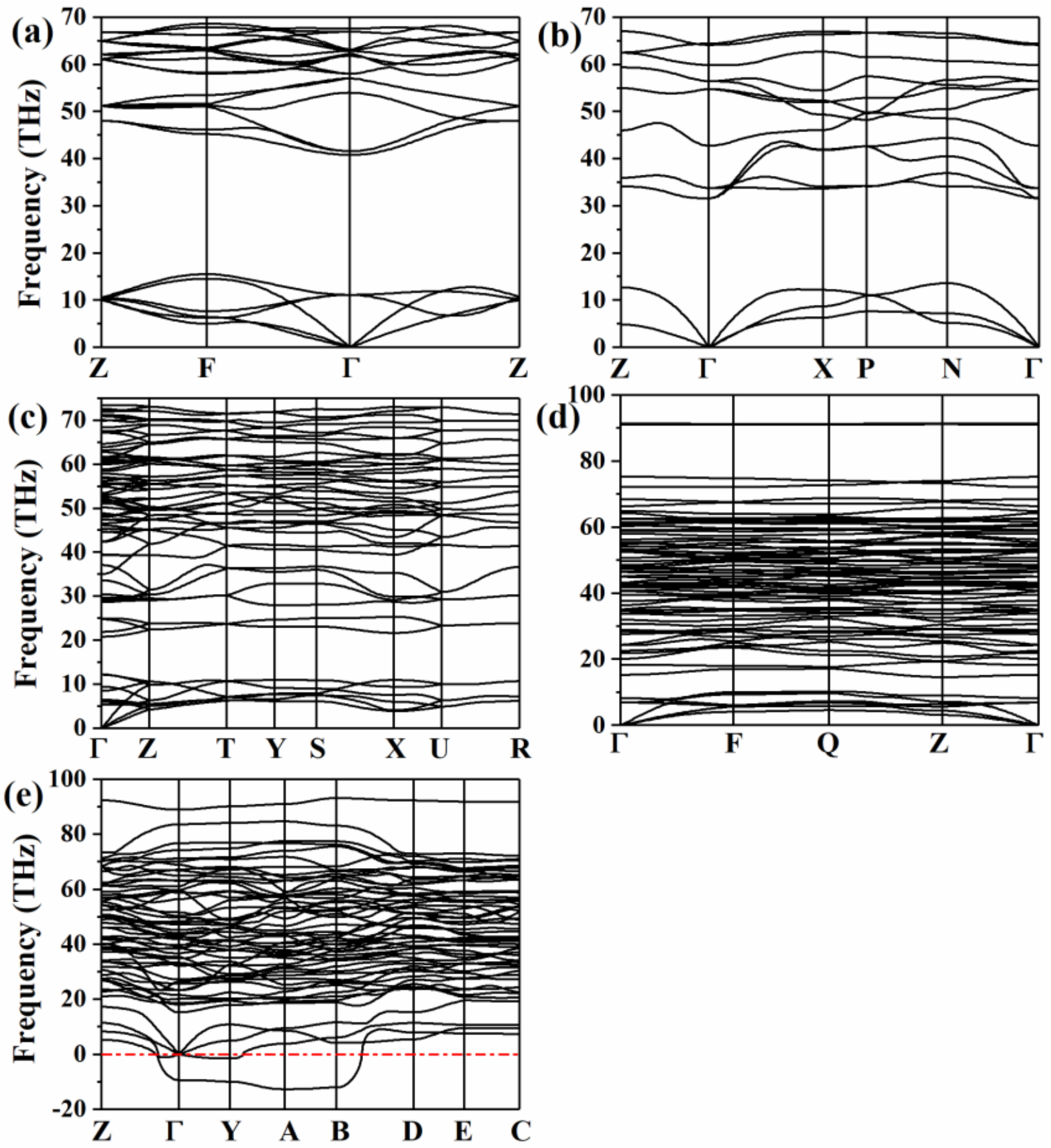

Fig. S17. The calculated phonon dispersion curves of (a) $R\bar{3}c$-$ZrH_3$, (b) $I4/mmm$-$ZrH_4$, (c) $Cmc2_1$-$ZrH_6$, (d) $P\bar{1}$-$ZrH_{10}$ and (e) $P2/m$-$ZrH_{17}$ at 300 GPa The absence of imaginary frequency in $ZrH_3$, $ZrH_4$, $ZrH_6$ and $ZrH_{10}$ suggests their dynamic stabilities, while $ZrH_{17}$ is unstable due to the appearance of negative phonon mode.

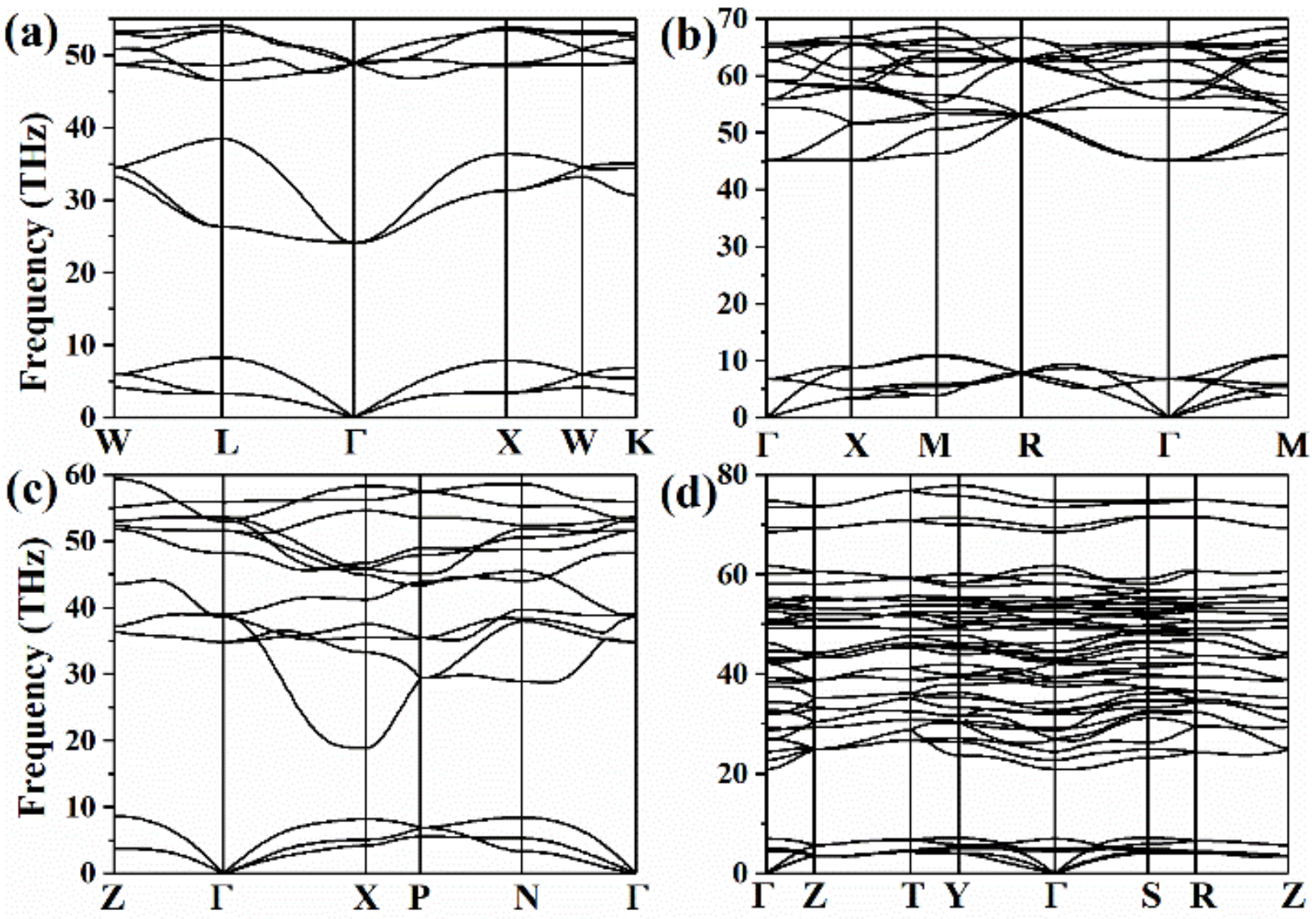

Fig. S18. The calculated phonon dispersion curves of (a) $Fm\bar{3}m$-$LuH_3$ at 100 GPa, (b) $Pm\bar{3}n$-$LuH_3$ at 300 GPa, (c) $I4/mmm$-$LuH_4$ at 200 GPa and (d) $Cmc2_1$-$LuH_7$ at 200 GPa. The absence of imaginary frequency of Lu hydrides confirms their dynamic stabilities.

# EOS for $MH_3$ (M=Hf, Zr and Lu)

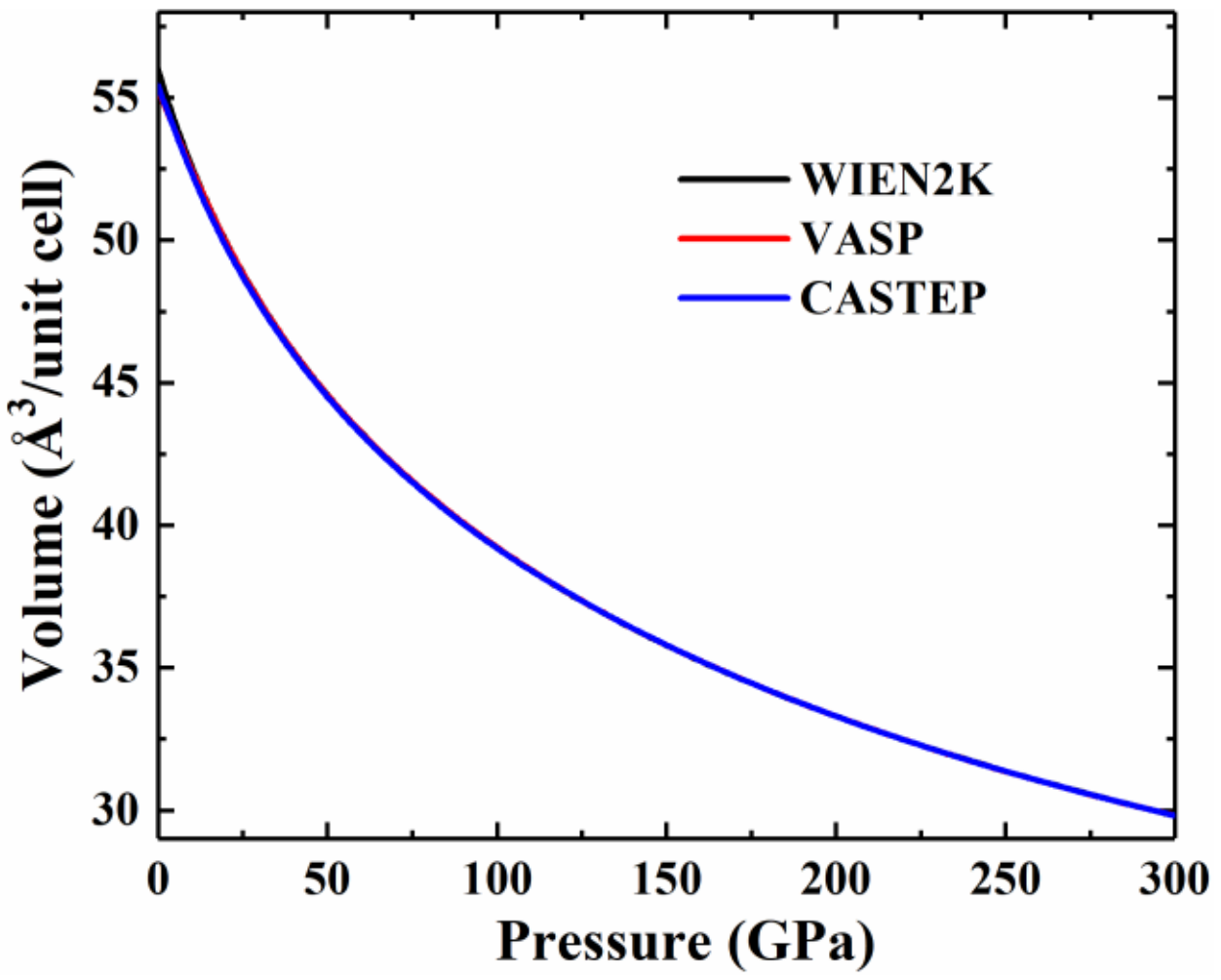

Fig. S20. Volumes as a function of pressures for $Pm\overline{3}n$-$HfH_3$ calculated by using PAW potential in VASP calculations, OTF potentials in CASTEP calculations and full-potential in WIEN2K calculations. The GGA_PBE exchange correlation functional was chosen in the three calculations, giving identical results. The purpose of this comparison is to test the validity of pseudopotentials in geometrical optimization and ZPE calculations below 300 GPa.

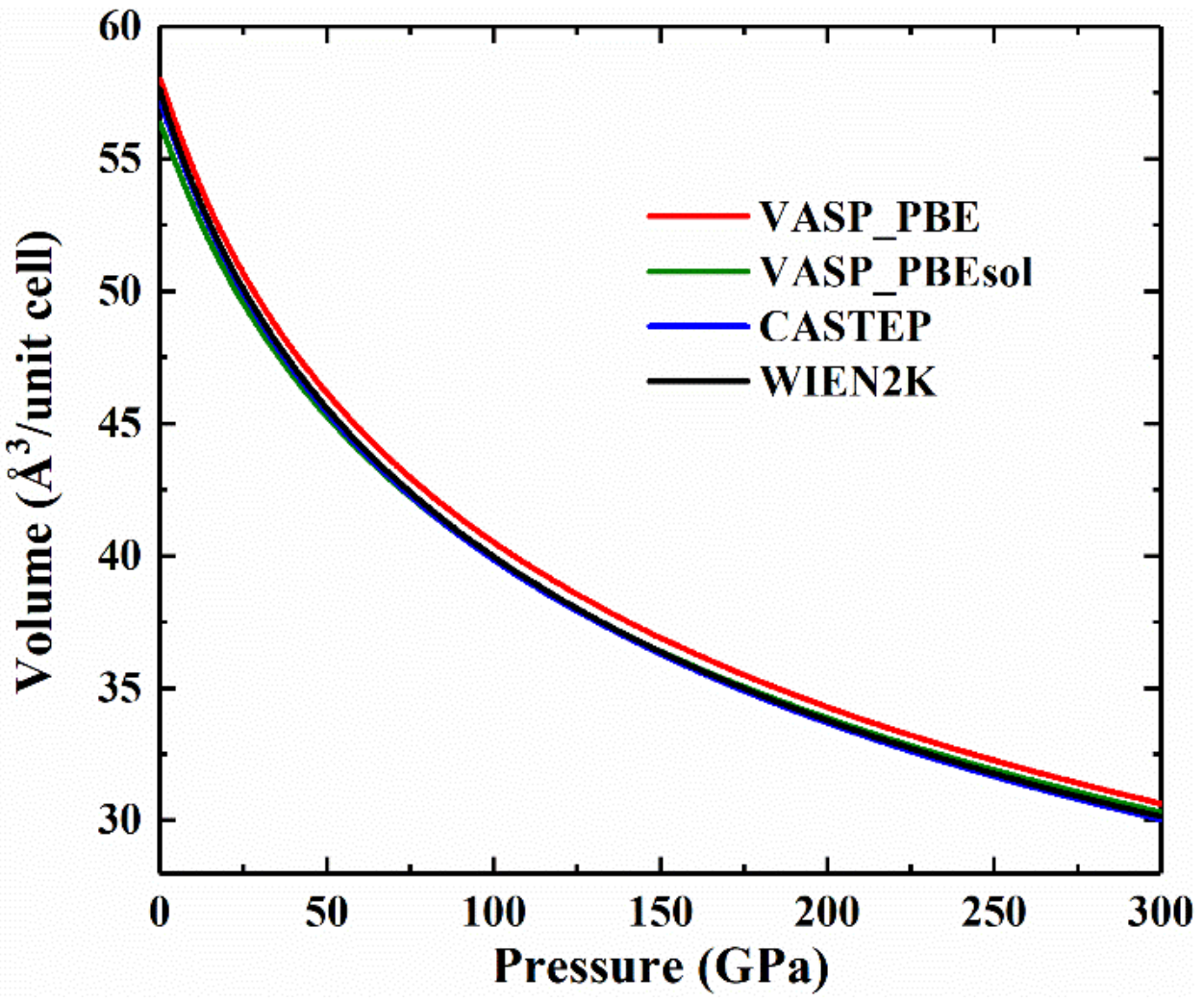

Fig. S21. Volumes as a function of pressures for $Pm\overline{3}n$-$ZrH_3$ calculated by using PAW potential (GGA_PBE and GGA_PBEsol exchange correlation functional) in VASP calculations, OTF potentials in CASTEP calculations and full-potential in WIEN2K. It is found that OTF potentials, PAW (GGA_PBEsol) potentials and full-potential calculations gave identical results, but VASP calculations with GGA_PBE exchange correlation functional deviate.

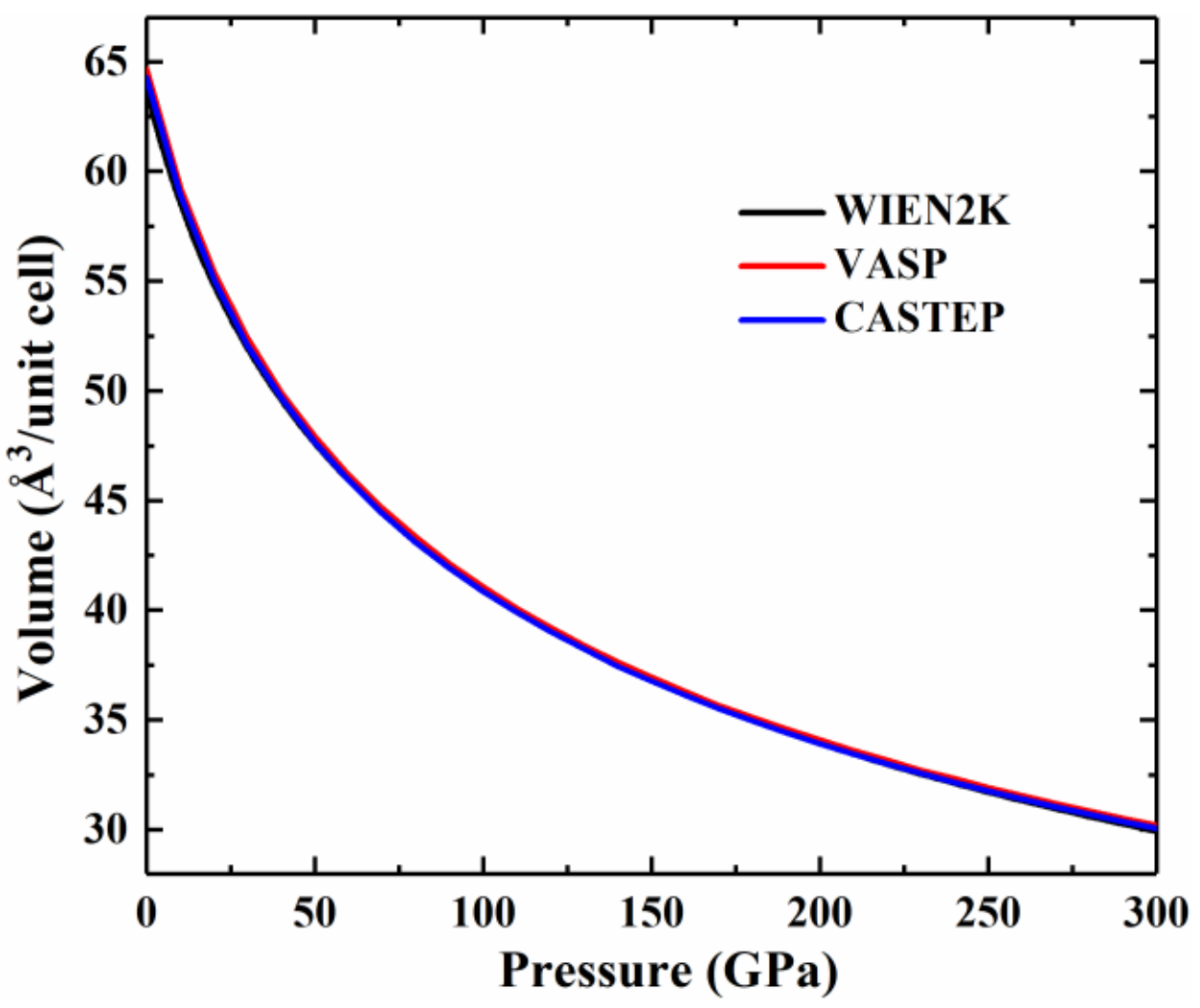

Fig. S22. Volumes as a function of pressures for $Pm\bar{3}n$-$LuH_3$ calculated by using PAW potential in VASP calculations, OTF potentials in CASTEP calculations and full-potential in WIEN2K calculations. The GGA_PBE exchange correlation functional was chosen in the three calculations, giving identical results. The purpose of this calculation is to test the validity of the PAW potentials adopted under pressures up to 300 GPa.

# Structural information

Table S7. Structural information of predicted hydrides.

| Space group<br>Pressure | Lattice parameters<br>(Å, °) | Atomic coordinates (fractional) | Sites |
|---|---|---|---|
| | a = 2.911 | H1 0.000 0.000 0.000 | 2a |
| $I4/mmm$-HfH | b = 2.911 | Hf1 0.000 0.000 0. 500 | 2b |
| 100 GPa | c = 3.676 | | |
| | α = β =γ = 90 | | |
| | a = 2.694 | H1 0.000 0.000 1.000 | 3a |
| $R\bar{3}m$-HfH | b = 2.694 | Hf1 0.000 0.000 0.500 | 3b |
| 300 GPa | c = 5.677 | | |
| | α = β =90 | | |
| | γ = 120 | | |
| | a = 3.222 | H1 0.500 0.000 0.250 | 4d |
| $I4/mmm$-$HfH_2$ | b = 3.222 | Hf1 0.000 0.000 0.000 | 2a |
| 100 GPa | c = 3.433 | | |
| | α = β = γ = 90 | | |
| | a = 3.368 | H1 -0.500 1.250 -2.000 | 6c |
| $Pm\bar{3}n$-$HfH_3$ | b = 3.368 | Hf1 0.000 0.000 -1.000 | 2a |
| 100 GPa | c = 3.368 | | |
| | α = β = γ = 90 | | |
| | a = 7.017 | H1 0.284 0.073 0.149 | 48e |
| $I\bar{4}3d$-$Hf_4H_{15}$ | b = 7.017 | H25 -0.125 -0.000 0.250 | 12b |
| 100 GPa | c= 7.017 | Hf1 0.043 0.043 0.043 | 16c |
| | α = β = γ = 90 | | |
| $Fddd$-$HfH_4$ | a = 10.180 | H1 0.171 0.085 1.764 | 32h |
| 200 GPa | b = 4.839 | Hf1 -0.250 0.750 1.250 | 8a |
| | c = 2.912 | | |
| | α = β = γ = 90 | | |
| | a =2.917 | H1 -0.217 -0.575 -0.018 | 8b |
| | b = 5.482 | H5 0.000 -0.641 -0.311 | 4a |
| $Cmc2_1$-$HfH_6$ | c = 4.818 | H6 0.000 -0.845 0.004 | 4a |
| 300 GPa | α = β = γ = 90 | H7 0.000 -0.896 -0.410 | 4a |
| | | H11 0.000 -0.008 -0.221 | 4a |
| | | Hf1 0.000 -0.322 -0.203 | 4a |
| | a=2.731 | H1 0.455 0.022 -0.187 | 2i |
| | b=2.831 | H2 -0.307 0.467 -0.058 | 2i |
| $P\bar{1}$-$HfH_{10}$ | c=3.626 | H3 -0.225 1.053 -0.952 | 2i |
| 300 GPa | α = 103.41 | H4 0.126 0.405 -0.838 | 2i |
| | β = 87.13 | H5 -0.321 0.568 -0.642 | 2i |
| | γ = 66.40 | Hf6 -0.000 0.000 -0.500 | 1b |
| $P6_3/mmc$-$HfH_{10}$ 300 GPa | a = 4.633<br>b = 4.633<br>c = 2.607<br>α = β = 90<br>γ = 120 | H1 0.375 0.077 0.250<br>H13 0.885 0.115 0.750<br>H19 0.000 0.000 0.250<br>Hf1 0.667 0.333 0.750 | 12j<br>6h<br>2b<br>2c |
| $P6_3/mmc$-$ZrH_{10}$ 300 GPa | a = 4.661<br>b = 4.661<br>c = 2.631<br>α = β = 90<br>γ = 120 | H1 0.370 0.076 0.250<br>H13 0.886 0.114 0.7500<br>H19 0.000 0.000 0.250<br>Zr1 0.667 0.333 0.750 | 12j<br>6h<br>2b<br>2c |
| $P6_3/mmc$-$ScH_{10}$ 250 GPa | a = 4.559<br>b = 4.559<br>c = 2.635<br>α = β = 90<br>γ = 120 | H1 0.384 0.080 0.250<br>H13 0.882 0.118 0.750<br>H19 0.000 0.000 0.250<br>Sc1 0.667 0.333 0.750 | 12j<br>6h<br>2b<br>2c |
| $P6_3/mmc$-$LuH_{10}$ 200 GPa | a = 4.806<br>b = 4.806<br>c = 2.784<br>α = β = 90<br>γ = 120 | H1 0.626 0.705 0.7500<br>H13 0.886 0.771 0.750<br>H19 0.000 0.000 0.250<br>Lu1 0.667 0.333 0.750 | 12j<br>6h<br>2b<br>2c |